\documentclass[12pt,letterpaper]{article}        
\usepackage{jheppub}
\pdfoutput=1\usepackage[utf8]{inputenc}
\usepackage{float}
\usepackage{amsmath,amssymb}
\usepackage{amsfonts}
\usepackage{slashed}
\usepackage{xcolor}
\usepackage{changes}

\begin{document}

% Use the \preprint command to place your local institutional report
% number in the upper righthand corner of the title page in preprint mode.
% Multiple \preprint commands are allowed.
% Use the 'preprintnumbers' class option to override journal defaults
% to display numbers if necessary
%\preprint{}

%Title of paper

\title{Gapless and gapped holographic phonons}
%\title{Larger font}
\author[1,2]{Andrea Amoretti,}
\author[3,4]{Daniel Are\'an,}
\author[4]{Blaise Gout\'{e}raux,} 
\author[5]{and Daniele Musso}

\emailAdd{andrea.amoretti@ge.infn.it}
\emailAdd{daniel.arean@uam.es}
\emailAdd{blaise.gouteraux@polytechnique.edu}
\emailAdd{daniele.musso@usc.es}

\preprint{CPHT-RR059.102019, IFT-UAM/CSIC-19-133}

\affiliation[1]{Dipartimento di Fisica, Universit\`a di Genova,
via Dodecaneso 33, I-16146, Genova, Italy and I.N.F.N. - Sezione di Genova
}
\affiliation[2]{Physique  Th\'{e}orique  et  Math\'{e}matique  and  International  Solvay  Institutes  Universit\'{e}  Libre de Bruxelles, C.P. 231, 1050 Brussels, Belgium}
\affiliation[3]{Instituto de F\'\i sica Te\'orica UAM/CSIC,
	Calle Nicol\'as Cabrera 13-15, Cantoblanco, 28049 Madrid, Spain}
\affiliation[4]{CPHT, CNRS, Ecole polytechnique, IP Paris, F-91128 Palaiseau, France}
\affiliation[5]{Universidade  de  Santiago  de  Compostela  (USC) and  Instituto  Galego  de  F\'{i}sica  de  Altas  Enerx\'{i}as  (IGFAE), Rúa de Xoaquín Díaz de Rábago, Campus Vida, 15705 Santiago de Compostela, A Coruña, Spain}

\abstract{We study a holographic model where translations are both spontaneously and explicitly broken, leading to the presence of (pseudo)-phonons in the spectrum.
 The weak explicit breaking is due to two independent mechanisms: a small source for the condensate itself and
additional linearly space-dependent marginal operators. The low energy dynamics of the model is described by Wigner crystal hydrodynamics. In absence of a source for the condensate, the phonons remain gapless, but momentum is relaxed.
Turning on a source for the condensate damps and pins the phonons.
 Finally, we verify that the universal relation between the phonon damping rate, mass and diffusivity reported in \cite{Amoretti:2018tzw} continues to hold in this model for weak enough explicit breaking.
}

\maketitle

\section{Introduction}
When translations are spontaneously broken in an otherwise translation invariant system, new gapless modes appear in the spectrum. 
These are the Nambu-Goldstone bosons of broken translations (which we will call \emph{phonons}, by a slight abuse of terminology). 
The low energy effective theory describing the dynamics around equilibrium at long wavelengths and late times is that of Wigner crystals, see \cite{chaikin_lubensky_1995} for a review. 
The phonons have a dramatic impact on the transverse spectrum: the transverse phonon mixes with transverse momentum to give two shear sound modes
\begin{equation}
\label{hydroappgapless}
\omega_{shear}=\pm\sqrt{\frac{G}{\chi_{PP}}}\,q-\frac{i}2 q^2\left(\xi_\perp+\frac{\eta}{\chi_{PP}}\right)+O(q^3)\,,
\end{equation}
 with a velocity proportional to the square root of the elastic shear modulus $G$
 and an attenuation controlled by the shear viscosity $\eta$ and the transverse phonon diffusivity $\xi_\perp$.  $\chi_{PP}$ is the momentum static susceptibility. In contrast, in a fluid, the only hydrodynamic mode in the transverse sector is purely diffusive
\begin{equation}
\label{disprelsheardiff}
\omega=-i \frac{\eta}{\chi_{PP}} q^2+O(q^4)\,.
\end{equation}
In other words, solids propagate shear sound waves, and fluids do not. 

When translations are explicitly broken (say by disorder or coupling to the underlying ionic lattice), the transverse sector of a fluid no longer contains any hydrodynamic mode. Instead, the shear diffusive mode \eqref{disprelsheardiff} is gapped 
\begin{equation}
\label{dispreldrude}
\omega=-i \Gamma- \frac{\eta}{\chi_{PP}} q^2+O(\Gamma^2,q^2\Gamma,q^4)\,,
\end{equation}
and encodes the relaxation of transverse momentum. At the longest times and distances $\omega,q\ll\Gamma$, the spectrum is completely incoherent. As $\Gamma$ is increased from zero, the longitudinal sound modes undergo a collision on the imaginary axis into a gapped, pseudo-diffusive mode (relaxation of longitudinal momentum) and a diffusive mode (thermal diffusion following from the conservation of energy).

Studying the long wavelength dynamics of holographic phases with explicitly broken translations has recently attracted a lot of attention and led to a thorough understanding of the physics of slow momentum relaxation (see \cite{Hartnoll:2016apf} for a review and references therein). While progress was also made for translations broken by inhomogeneous sources \cite{Horowitz:2012ky,Chesler:2013qla,Hartnoll:2014cua,Donos:2014yya,Donos:2017ihe}, the brunt of the effort exploited so-called homogeneous models \cite{Donos:2012js,Vegh:2013sk,Donos:2013eha,Andrade:2013gsa,Gouteraux:2014hca,Baggioli:2014roa,Donos:2014oha}, whereby a global bulk symmetry is broken together with translations. This leads to ordinary differential equations in the bulk rather than partial differential equations, which is a significant technical simplification. 
In spite of their simplicity, these models accurately capture the dynamics of slow momentum relaxation when translations are weakly broken \cite{Davison:2013jba,Davison:2014lua,Davison:2015bea,Andrade:2018}.\footnote{These holographic homogeneous models inspired analogous constructions \cite{Musso:2018wbv}
which account for the translation-breaking dynamics in a purely field-theoretical context.
For a generalization of such constructions to inhomogeneous case see \cite{Musso:2019kii}.}

In a two-dimensional, isotropic Wigner crystal, explicit breaking of translations gaps the transverse sound modes, relaxing both the transverse momentum and the transverse phonon \cite{Delacretaz:2017zxd}:
\begin{equation}
\label{wchydrogapped}
\omega=-\frac{i}{2}\left(\Gamma+\Omega\right)\pm\frac12\sqrt{4\omega_o^2-(\Gamma-\Omega)^2}+O(q)\,.
\end{equation}
Here $\Omega$ is the phonon damping rate and $\omega_o$ the pinning frequency, related to the phonon mass by $\omega_o^2=G m^2/\chi_{PP}$.\footnote{The pinning frequency can also be written more intuitively as $\omega_o=c_\perp k_o$ with $k_o\equiv m$ the inverse length scale defined by the phonon mass and $c_\perp\equiv \sqrt{G/\chi_{PP}}$ the shear sound velocity.}
The same occurs in the longitudinal sector for the longitudinal momentum and longitudinal phonon, leaving a single, diffusive hydrodynamic mode, encoding the conservation of energy.

Pinning of density waves by a soft explicit breaking of translations introduces two new relaxation parameters, the phonon mass $m$ and damping rate $\Omega$ \cite{Delacretaz:2016ivq,Delacretaz:2017zxd}. In \cite{Delacretaz:2016ivq,Delacretaz:2017zxd}, only defects were considered as a microscopic origin for the damping rate $\Omega$. However, as demonstrated in \cite{Amoretti:2018tzw} and further investigated in \cite{Ammon:2019wci,Baggioli:2019abx}, damping by explicit breaking of translations enters in the low energy effective theory in precisely the same way. 

Spontaneous breaking of translations was also studied holographically starting with \cite{Ooguri:2010kt,Donos:2011bh}, leading to the construction of fully backreacted, inhomogeneous, spatially modulated phases \cite{Donos:2013gda,Withers:2013loa,Withers:2014sja,Ling:2014saa,Jokela:2014dba,Cremonini:2016rbd,Cai:2017qdz,Andrade:2017ghg}. While homogeneous models with a Bianchi VII global symmetry where studied in parallel (see eg \cite{Donos:2012wi}), this setup remains less explored due to its higher technical difficulty. Simpler homogeneous models were investigated in \cite{Amoretti:2016bxs,Alberte:2017oqx,Amoretti:2017frz,Amoretti:2017axe}. This opened the way to extensive studies of the effective low energy dynamics of holographic density waves, pinned or not, \cite{Jokela:2017ltu,Andrade:2017cnc,Alberte:2017cch,Amoretti:2018tzw,Andrade:2018,Donos:2019tmo,Ammon:2019wci,Amoretti:2019cef,Ammon:2019apj,Baggioli:2019abx,Donos:2019hpp}. The effective, hydrodynamic theory of pinned charge density waves of \cite{Delacretaz:2016ivq,Delacretaz:2017zxd} provides in most cases the correct framework to interpret these results.\footnote{As explained above, the longitudinal diffusive mode reported in \cite{Ammon:2019wci,Baggioli:2019abx} corresponds to diffusion of energy \cite{Delacretaz:2017zxd}. The other two gapped modes are the gapped longitudinal phonon and momentum density. The case of \cite{Jokela:2017ltu} is conceptually distinct, as momentum is decoupled. 
\cite{Donos:2019tmo,Donos:2019hpp} work in the limit where the momentum relaxation rate is much larger than phonon damping and pinning. 
\cite{Ammon:2019apj} finds a discrepancy between the hydrodynamic prediction in the longitudinal sector and the holographic result, which is not resolved at this point. {It is likely though that a resolution lies in a proper accounting of the background strain of these phases, along the lines explained in \cite{Armas:2019sbe}. This sources new contributions in the dispersion relations of the longitudinal modes.}}

In \cite{Amoretti:2016bxs,Amoretti:2017frz,Amoretti:2017axe,Amoretti:2018tzw,Amoretti:2019cef}, we studied a bi-dimensional holographic `Q-lattice' \cite{Donos:2013eha}, where the UV CFT is deformed by complex, neutral scalar operators whose phases are linear in the boundary spatial coordinates and break translations. If the bulk field dual to the moduli of the complex operators is not sourced, $\lambda=0$, the breaking is spontaneous. The complex scalar deformations are the order parameters.\footnote{As described in \cite{Amoretti:2017frz,Amoretti:2017axe}, a UV analysis shows that the holographic setup we consider corresponds to a complex scalar deformation for each spatial dimension where 
the moduli of all the operators are constrained to be equal. 
}
In \cite{Amoretti:2019cef}, we verified that transverse hydrodynamic modes with the dispersion relation \eqref{hydroappgapless} are present in the spectrum. In \cite{Amoretti:2018tzw}, we showed that these modes become gapped if a small source is turned on, with a dispersion relation \eqref{wchydrogapped}.
We found that $\Omega\simeq G m^2\Xi$, where $\Xi\equiv\xi_{\parallel}/(K+G)$ is defined from the bulk and shear moduli $K$, $G$ and the longitudinal phonon diffusivity $\xi_{\parallel}$.\footnote{For an isotropic crystal, $\xi_{\parallel}/(K+G)=\xi_\perp/G\equiv\Xi$, so this relation can also be expressed in terms of the transverse phonon diffusivity $\xi_\perp$.}\footnote{This relation was later confirmed in a holographic massive gravity model \cite{Ammon:2019wci}, and in \cite{Donos:2019txg} in the limit where only the global symmetry is broken.} In other words, the ratio $\Omega/m^2$, where both quantities depend on the microscopic details of the system and of the breaking of translations, is given by universal thermodynamic and hydrodynamic data.

In this work, we study the interplay between two conceptually distinct sources of relaxation, as described in section \ref{section:holomodel}. 
{One is the source $\lambda$ for the Q-lattice, which can be thought of as a source for the translation-breaking order parameters themselves}. 
The other comes from massless scalar fields linear in the boundary spatial coordinates, in the spirit of \cite{Andrade:2013gsa}. They are dual to a marginal deformations of the UV CFT, with a source $\ell\neq0$. These deformations break translations weakly, relax momentum and lead to finite DC conductivities. They leave intact the global U(1) symmetries of the Q-lattice, which shift the phases of the complex scalars by a constant. As a consequence, setting $\lambda=0$, the spectrum still contains gapless modes even though momentum relaxes \cite{Donos:2019tmo}. 

In section \ref{section:lambda0}, we study the spectrum in the transverse sector with only the source $\ell$ turned on.
At zero wavevector $q=0$, we find a single gapped mode with dispersion relation $\omega=-i\Gamma$, in marked contrast to \eqref{wchydrogapped}. {This tension is resolved by observing that when $\ell\neq0$ and $\lambda=0$, the phonons remain gapless and only momentum relaxes. Thus we should set $m=\Omega=0$ in \eqref{wchydrogapped}.
For small $\ell\ll\mu$, the location of the remaining gapped pole agrees with the memory matrix prediction for the momentum relaxation rate of the system \eqref{GMM}. }
At small nonzero wavector $q\neq0$, we find two light modes (one gapless, the other gapped) with dispersion relation \eqref{hydrodispersiontran}, 
which are the modes predicted by Wigner crystal hydrodynamics in the absence of any phonon mass or damping $m=\Omega=0$, 
but with nonzero momentum relaxation $\Gamma\neq0$. 
The comparison between the holographic and hydrodynamic dispersion relations allows us to provide a prescription to compute unambiguously the phonon shear modulus $G$ in the holographic model, by subtracting from the quantity $-\lim_{\omega\to0}\textrm{Re}\left[G^R_{T_{xy}T_{xy}}\right]$ a negative contribution proportional to $\ell^2$. Using this prescription for $G$, we find that the transverse quasinormal modes (QNMs) computed holographically agree very well with the hydrodynamic dispersion relations.

With this prescription in hand, we then turn on both sources of explicit translation breaking $\lambda\neq0$, $\ell\neq0$ in section \ref{section:ellandlambdanonzero}. 
At zero wavector and for small sources $\lambda,\ell\ll\mu$, the spectrum contains two light, gapped modes with a dispersion relation given by \eqref{wchydrogapped}. In contrast to \cite{Amoretti:2018tzw}, $\Gamma$ is non-negligible with $\ell\neq0$, which allows us to further test the validity of the hydrodynamic theory of relaxed Wigner crystals.
We also check that the hydrodynamic prediction \eqref{conduacfinitelambda} for the ac conductivity agrees with our numerics. 

Next, we test the validity of the relation $\Omega \simeq G m^2 \Xi$ at varying $\lambda$ and $\ell$. At high temperature, we find that it is quite insensitive to the value of the source $\ell$. This is in agreement with our earlier result \cite{Amoretti:2018tzw} that the phonon mass and damping rate are controlled by $\lambda$ at leading order (see also \cite{Donos:2019txg,Donos:2019hpp}). At low temperatures, the system quickly becomes more incoherent and relaxation is stronger, which leads to a failure of the relation above as $\ell$ increases, see figure \ref{comparizonmxifig}.

Finally, at nonzero wavevector, the dispersion relation of the QNMs matches the hydrodynamic dispersion relation \eqref{hydrodiffusivefl}.

We conclude with some final comments and future directions in section \ref{sec:outlook}.
\\
\vskip0.5cm
{\bf Note added:} While this project was underway, \cite{Donos:2019hpp} appeared which contains some overlap with our results. The analysis of \cite{Donos:2019hpp} focuses on the longitudinal sector, at scales $\omega\sim\Omega\sim\omega_o\ll\Gamma$. Where applicable, their results appear compatible with ours.

\section{The holographic model}
\label{section:holomodel}
We consider the following holographic model:
\begin{equation}\label{action}
S=\int d^{4}x\,\sqrt{-g}\left[R-\frac{1}{2}\partial\phi^2-\frac{1}{4}Z(\phi)F^2+V(\phi)-\frac{1}{2}Y(\phi)\sum_{I=1}^{2}\partial \psi_I^2-\frac{1}{2}\tilde Y(\phi)\sum_{I=1}^{2}\partial\tilde \psi_I^2\right]  .
\end{equation}
A similar model has been previously studied in \cite{Donos:2019tmo}. 
Choosing the following Ansatz for the scalars $\psi_I$ and $\tilde{\psi}_I$
\begin{equation}\label{backgroundansatzpsi}
\psi_I= k \delta_{Ij} x_j \ , \qquad \tilde{\psi}_I=\ell \delta_{Ij} x_j \ , \qquad \text{ with } x_j=\{x,y\} \ ,
\end{equation}
breaks translations along the spatial directions $x$ and $y$ of the dual field theory. For simplicity, we restrict to an isotropic breaking and choose each scalar in the pairs $\psi_I$, $\tilde\psi_I$ to be aligned along one spatial direction only, eg $\psi_1=k x$.\footnote{{This is an Ansatz, and does not guarantee that other solutions do not exist, eg inhomogeneous solutions with a lower free energy. This is an interesting question for future work.}} So in the following, we no longer distinguish between capital and lowercase latin indices. 
The Ansatz \eqref{backgroundansatzpsi} breaks both translations and the global symmetry of the scalars $\psi_I\mapsto\psi_I+c_I$, $\tilde\psi_I\mapsto\tilde\psi_I+\tilde c_I$ to a diagonal $U(1)$ combination. Thus translations are broken homogeneously, and the remaining $U(1)$ allows the background metric, scalar field $\phi$ and gauge potential $A_t$ to depend only on the radial coordinate $r$ \cite{Donos:2013eha,Andrade:2013gsa}.
This is a considerable simplification.

We are interested in solutions with Anti de Sitter asymptotic boundary conditions. To this end, we choose the boundary $\phi\to0$ behavior of the scalar couplings as:
\begin{equation}\label{uvphipotentials}
\begin{split}
&V_{uv}(\phi)=-6-\phi^2+\mathcal{O}(\phi^3) \ , \; \; Z_{uv}(\phi)=1+\mathcal{O}(\phi) \ , \; \;\\
&  Y_{uv}(\phi)=Y_2 \phi^2+ \mathcal{O}(\phi^4) \ , \; \; \tilde{Y}_{uv}(\phi)=\tilde{Y}_0+ \mathcal{O}(\phi^2) \ .
\end{split}
\end{equation}
Near the AdS boundary $r\to0$, the scalar field $\phi$ behaves as:
\begin{equation}\label{scalaruvexp}
\phi= \lambda r+ \phi_{(v)}r^2+ \mathcal{O}(r^3) \ ,
\end{equation}
so that indeed $\phi\to0$ at the boundary.

 As explained in \cite{Amoretti:2017frz,Amoretti:2017axe,Donos:2019tmo}, when the leading mode $\lambda$ of the scalar field $\phi$ is set to zero, the operators $\psi_I$ break translations spontaneously in the dual field theory. In brief, this is because the small $\phi$ asymptotics of the coupling $Y$ allows to rewrite the scalars $(\phi,\psi_I)$ into a pair of complex scalars $\Phi\sim\phi \exp(i\psi_I)$, as in \cite{Donos:2013eha}. If $\lambda=0$, the background solution for this scalar breaks translations spontaneously in the dual field theory, while it breaks them explicitly if $\lambda\neq0$. {In past work \cite{Amoretti:2017frz,Amoretti:2017axe,Amoretti:2018tzw,Amoretti:2019cef}, we have explicitly constructed both types of solutions wih $\lambda=0$ or $\lambda\neq0$ in the model \eqref{action}, either with the choice of potentials \eqref{uvphipotentials} or different potentials.}
 
 In principle, thermodynamically stable phases breaking translations spontaneously minimize the free energy with respect to $k$ \cite{Donos:2013eha}. 
 This is indeed the logic followed in previous works to construct the backreacted, inhomogeneous solutions \cite{Donos:2013wia,Withers:2013loa}. The solutions constructed in the present work do not minimize the free energy with respect to $k$ (or would do so at $k=0$, in which case translations are not broken at all). Yet, there is by now abundant evidence \cite{Amoretti:2017frz,Amoretti:2017axe,Amoretti:2018tzw,Amoretti:2019cef,Donos:2019hpp,Donos:2019tmo} that the low energy dynamics of such phases is still well described by Wigner crystal hydrodynamics. As we showed in \cite{Amoretti:2017frz}, stable phases can be found by including eg higher derivative terms in the bulk action, with generally small values for their couplings. It is likely a Chern-Simons term would also work, as in inhomogeneous constructions.

 The field $\tilde{\psi}_I$ with $\ell\neq0$ always breaks translations explicitly due to the different boundary behavior of $\tilde Y$, as in \cite{Andrade:2013gsa}. 
 
 Our main goal is to identify the effective field theory described by this holographic model. Once the UV behavior \eqref{uvphipotentials} is fixed, our results will be mostly independent on the specific choice of the potentials, but, for concreteness, the numeric results will be obtained for:
\begin{equation}
\begin{split}
&V(\phi)=-6 \cosh(\phi) \ , \; \; Z(\phi)=\cosh( 3 \phi)^{\frac{4}{3}} \ , \; \;\\
& Y(\phi)=12 \sinh(\phi)^2 \ , \; \; \tilde{Y}(\phi)= 12(1+\sinh(\phi)^2) \ .
\end{split}
\end{equation}

%%%%%%%%%%%%%%%%%%%%%%%%%%
\subsection{Background geometry and thermodynamics}
We are going to consider the following gauge for the metric:
\begin{equation}
ds^2=\frac{1}{r^2} \left(-U(r) dt^2+\frac{dr^2}{U(r)}+c(r)(dx^2+dy^2) \right) \ .
\end{equation}
The asymptotic expansion of the background fields near the boundary $r=0$, reads:
\begin{equation}
\begin{split}
&U(r)=1-\frac{3}{4}(8 \ell^2+\lambda^2) r^2+u_3 r^3+ \mathcal{O}(r^4) \ ,\\
&c(r)=1-\frac{3 \lambda^2}{4}r^2-\lambda\phi_{(v)} r^3+\mathcal{O}(r^4) \ ,\\
&A(r)=\mu-\rho r+\mathcal{O}(r^2) \ ,\\
&\phi(r)=\lambda r+\phi_{(v)}r^2+ \mathcal{O}(r^3) \ .
\end{split}
\end{equation}
There is a radially conserved quantity defined by the relation:
\begin{equation}
\label{radiallycons}
\left[-\rho A(r)+\frac{c(r)}{r^2}\left(\frac{U(r)}{c(r)}\right)'-k^2 \int_{r_h}^r  \frac{Y(\phi)}{r^2}dr-l^2 \int_{r_h}^r  \frac{\tilde{Y}(\phi)}{r^2}dr \right]'=0\,.
\end{equation}

We are interested in studying finite temperature states, which correspond to introducing a black hole with a regular horizon at $r=r_h$ in the bulk gravitational theory. The background fields have the following near horizon behavior:
\begin{equation}\label{horizonexp}
\begin{split}
&ds^2=-4 \pi T(r_h-r) dt^2+\frac{dt^2}{4 \pi T (r_h-r)}+\frac{s}{4 \pi}(dx^2+dy^2) +... \\ 
&A_t=A_h(r_h-r)+... \ , \qquad \phi=\phi_h+... \ ,
\end{split}
\end{equation} 
where $T$ and $s$ are the temperature and the entropy of the black hole respectively, namely:
\begin{equation}
s= 4 \pi \frac{c(r_h)}{r_h^2} \ , \qquad T=\frac{1}{4 \pi} \left.\sqrt{U'(r)^2-\frac{4U(r)}{r^2}}\right|_{r=r_h} \ .
\end{equation}

Evaluating \eqref{radiallycons} at the horizon and at the boundary, we obtain the relation:
\begin{equation}
\label{radiallycons2}
s T+ \mu \rho= -3 u_3+\lambda \phi_{(v)}-k^2 I_Y-\ell^2 I_{\tilde{Y}} \ ,
\end{equation}
with
\begin{equation}\label{renintegral}
I_Y=\int_0^{r_h}\frac{Y(\phi)}{r^2} dr \ ,\qquad I_{\tilde{Y}}=\lim_{\epsilon \rightarrow 0} \left[\int_{\epsilon}^{r_h}\frac{\tilde{Y}(\phi)}{r^2}dr-\frac{\tilde{Y}(0)}{\epsilon} \right] \ ,
\end{equation}
where in the last equality the UV divergence of $\int\tilde{Y}(r)/r^2$ is regulated by the second term.

Applying the usual holographic renormalization procedure \cite{deHaro:2000vlm}, one can compute the on-shell background action and the one-point functions of the model:
\begin{equation}
\label{1pointfunctions}
\begin{split}
\langle T^{tt} \rangle=&\epsilon=-{2u_3}+\lambda \phi_{(v)} \ , \; \; \langle T^{xx} \rangle=\langle T^{yy} \rangle=-{u_3}=p+k^2 I_Y+\ell^2 I_{\tilde{Y}}\\
&\qquad \; \; \langle J^t \rangle=\rho \ , \qquad \langle \mathcal{O}_{\phi} \rangle= \phi_{(v)} \ , 
\end{split}
\end{equation}
where $\epsilon$, $p$ and $\rho$ are the energy density, the pressure (identified from the renormalized on-shell action) and the charge density respectively.

Plugging these expressions in \eqref{radiallycons2}, we obtain the usual Smarr relation:
\begin{equation}
sT+\mu\rho=\epsilon+p\,.
\end{equation}

As explained in \cite{Amoretti:2017frz}, the momentum susceptibility $\chi_{PP}$ can be computed exactly in $k$ and $l$ by boosting the expressions \eqref{1pointfunctions} for the one-point functions of the various components of the stress energy tensor. This leads to:
\begin{equation}\label{chipp}
\chi_{PP}=s T+ \mu \rho +k^2 I_Y+\ell^2 I_{\tilde{Y}}=-3 u_3+\lambda \phi_{(v)} \ .
\end{equation}

In the rest of this work, we will be interested in the spectrum of transverse fluctuations at nonzero frequency and nonzero wavevector. Due to the homogeneity of the background solution, we can decompose them in plane waves as
\begin{equation}
\begin{split}
&\delta \psi_x=\delta \psi_x(r)e^{-i\omega t+i q y}\,,\quad \delta \tilde \psi_x=\delta \tilde \psi_x(r)e^{-i\omega t+i q y}\,,\\
& \delta a_x=\delta a_x(r)e^{-i\omega t+i q y}\,,\quad \delta g^x_{t}=\delta g_t^{x}(r)e^{-i\omega t+i q y}\,.
\end{split}
\end{equation}
For the computation of the quasinormal modes we use the determinant method following \cite{Kaminski:2009dh}.

%%%%%%%%%%%%%%%%%%
\section{Gapless phonons}
\label{section:lambda0}

We will not review Wigner crystal hydrodynamics in this work. However, the reader unfamiliar with this material is invited to consult \cite{Delacretaz:2017zxd} from which all dispersion relations quoted here can be derived, or \cite{Amoretti:2018tzw,Amoretti:2019cef,Ammon:2019apj} for more recent reviews of the formalism.

%%%%%%%%%%%%%%%%%%%
\subsection{Spectrum and conductivity at zero wavevector}
\label{q_zero}

\begin{figure}[H]
\begin{center}
\includegraphics[width=0.49\textwidth]{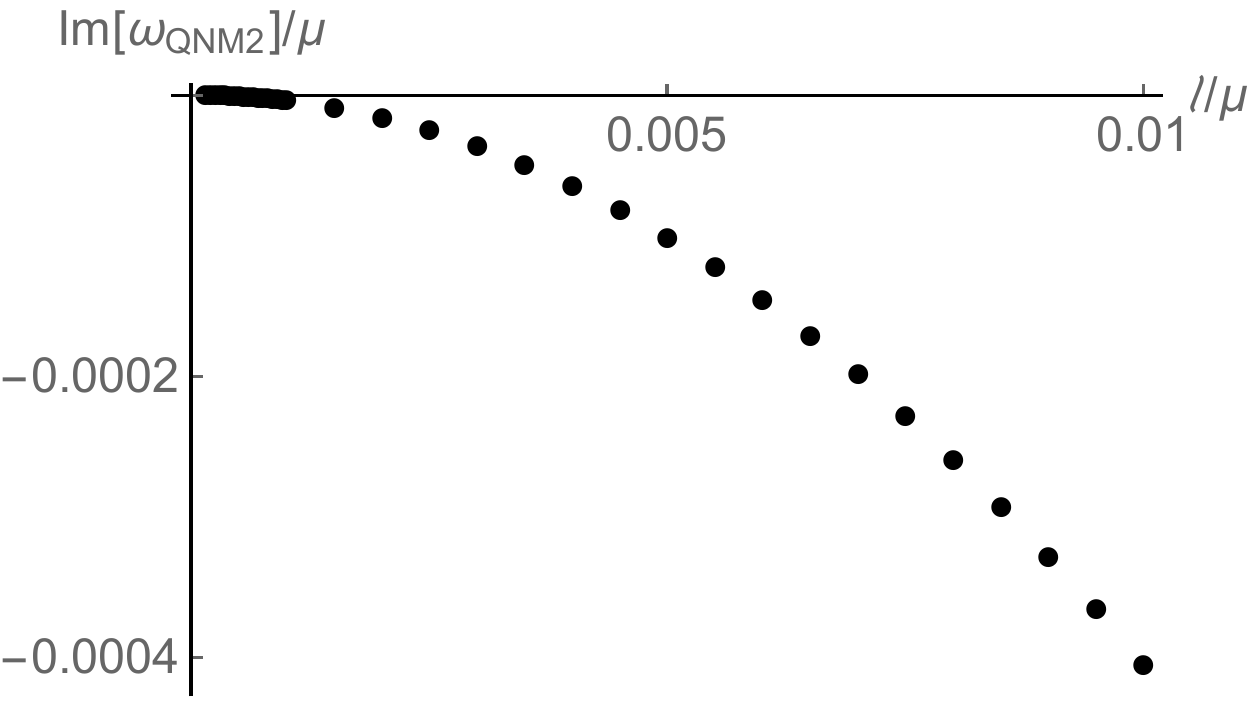}
\includegraphics[width=0.49\textwidth]{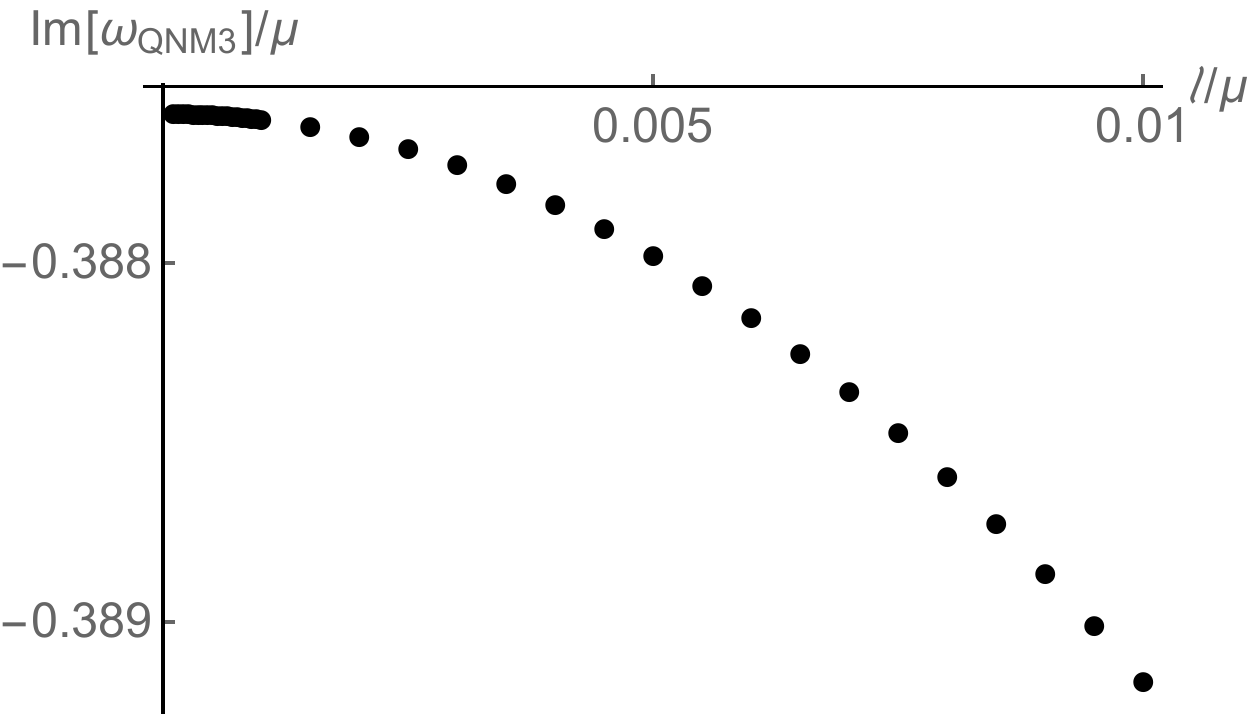}
\end{center}
\caption{Second and third purely imaginary QNMs for $({T}/{\mu},k/\mu,\lambda)=(0.1,0.01,0)$.
The first QNM (not depicted) stays fixed at the origin as ${\ell}/{\mu}$ is varied.
Left: QNM$_2$ is gapped for $\ell\neq0$ and becomes gapless for ${\ell}/{\mu}\to 0$.
Right: QNM$_3$ remains gapped as ${\ell}/{\mu}\to 0$. }
\label{QNM23}
\end{figure}

Setting $\lambda=0$, the background Ansatz of the fields $\psi$ breaks translations spontaneously, as explained above.
On the other hand, the background Ansatz of  the fields $\tilde \psi$ breaks translations explicitly.

In the absence of the fields $\tilde\psi$, the spectrum of modes includes gapless modes corresponding to the phonons and momentum densities along each spatial direction. 
As the global symmetry associated to the $\psi$ fields is left unbroken when $\ell\neq0$, 
we expect that only the momentum modes are gapped, and the phonons remain gapless. As we shall now see, this is exactly what happens. 

We have computed the lowest-lying QNMs for $T/\mu=0.1$ and $k/\mu=0.01$ while varying $\ell/\mu$ from $0$ to $0.01$.
We find a first quasinormal mode, QNM$_1$, at the origin, whereas the second and the third are purely imaginary 
(they are denoted by QNM$_{2,3}$ in figure \ref{QNM23}). 
QNM$_2$ becomes gapless as ${\ell}/{\mu}\to 0$ and at nonzero $\ell\ll\mu$, its gap is the momentum relaxation rate $\Gamma$. 
Instead, QNM$_3$ remains  gapped in the limit ${\ell}/{\mu}\to 0$. At frequencies comparable to its gap, we expect the effective description in terms of relaxed Wigner crystal hydrodynamics to break down.

To confirm that the gap of QNM$_2$ gives the momentum relaxation rate of the system, 
we compare it to the memory matrix prediction $\Gamma_{\text{MM}}$ (see eg \cite{Hartnoll:2016apf})
\begin{equation}\label{MMcom}
 \Gamma_{\text{MM}} = \frac{1}{\chi_{PP}} \lim_{\omega\to 0} \lim_{\ell\to 0} \frac{1}{\omega} \text{Im} G^R_{\partial_t P\partial_t P}(\omega,q=0)\ ,
\end{equation}
where $P$ is the momentum operator and $\chi_{PP}$ is the static momentum susceptibility computed in \eqref{chipp}.

The correlator in \eqref{MMcom} can be evaluated holographically relying on the Ward-Takahashi identity for the vev of the momentum density
$\partial_t \pi = - \ell\, \langle \delta O_{\tilde \psi}\rangle$
(valid at the level of bulk fluctuations). Thus,
\begin{equation}\label{GMM}
 \Gamma_{\text{MM}} = \frac{\ell^2}{\chi_{PP}} \lim_{\omega\to 0} \left. \frac{1}{\omega} \text{Im} G^R_{O_{\tilde \psi} O_{\tilde \psi}}(\omega,q=0)\right|_{\ell=0}
 = \frac{\ell^2 c_h \tilde Y(\phi_h)}{\chi_{PP} r_h^2} \ .
\end{equation}
Because of the zero relaxation limit in \eqref{MMcom}, one only needs to know the holographic correlator at $\ell=0$, hence the computations 
can be done as in the appendix of \cite{Amoretti:2018tzw}.
In figure \ref{GammaGamma}, we show that the absolute value of the imaginary part of QNM$_2$ is well approximated by $\Gamma_{MM}$.
Therefore, the QNM spectrum at $q=0$ confirms that momentum is relaxed (QNM$_2$) 
while the phonon (QNM$_1$) is neither damped nor pinned. 

\begin{figure}
\begin{center}
\includegraphics[width=0.49\textwidth]{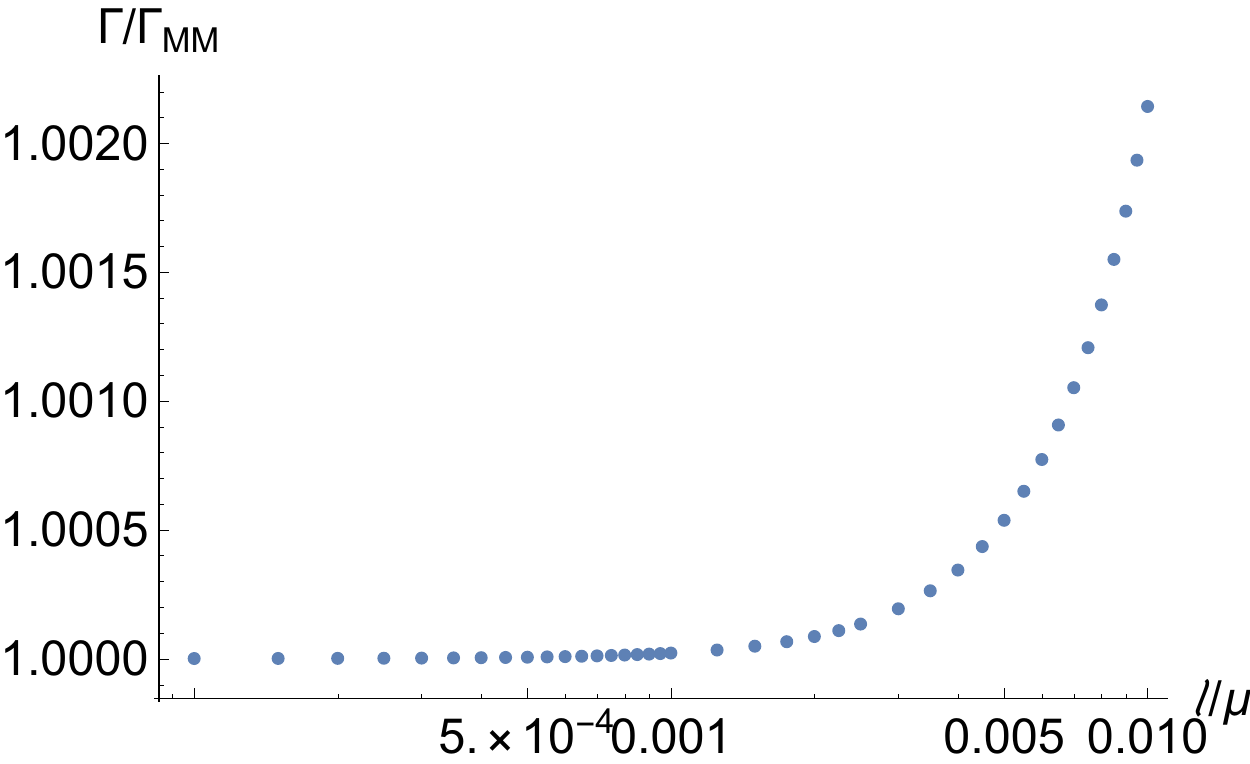}
\includegraphics[width=0.49\textwidth]{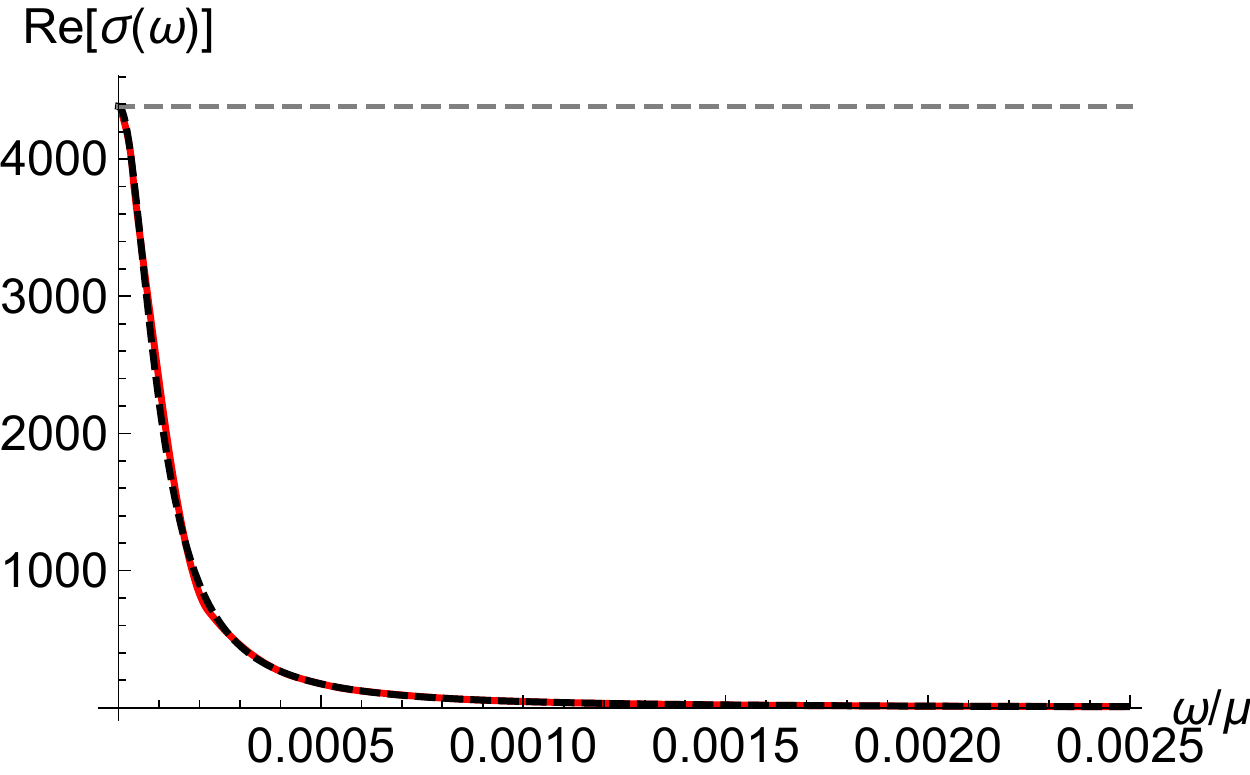}
\end{center}
\caption{In both panels, $({T}/{\mu},k/\mu,\lambda/\mu)=(0.1,0.01,0)$. Left: Comparison between $\Gamma_{\text{MM}}$ in \eqref{GMM} and $\Gamma$,  the gap of the purely imaginary QNM$_2$.
Right: Comparison at $\ell/\mu=5\times 10^{-3}$ between $\sigma_{\text{hydro}}(\omega)$ in \eqref{sig_hyd} (dashed, black) and the numerical $\sigma(\omega)$ (red); the dashed, gray line is the dc conductivity computed from the formula \eqref{sigmadc}.}
\label{GammaGamma}
\end{figure}

Adopting the notation of \cite{Delacretaz:2017zxd}, we have $\Gamma = \Gamma_{\text{MM}}$ and $\Omega=\omega_o=0$.
The hydrodynamic conductivity is
\begin{equation}\label{sig_hyd}
 \sigma_{\text{hydro}}(\omega) = \sigma_o + \frac{\rho^2}{\chi_{PP}} \frac{1}{\Gamma - i \omega}\ .
\end{equation}
Using for $\rho$ and $\chi_{PP}$ the results obtained in section \ref{section:holomodel}, as well as the  formula for $\sigma_o$ derived in \cite{Amoretti:2017frz},
\begin{equation}\label{sigmainc}
\sigma_o=\frac{1}{\chi_{PP}^2}\left[(sT+k^2 I_Y)^2Z(\phi_h)+\frac{4 \pi k^2I_Y^2 \rho^2}{s Y(\phi_h)}\right]\ ,
\end{equation}
we find excellent agreement with the ac conductivity computed numerically, see figure \ref{GammaGamma}, right plot.
The small frequency limit of the numerical ac conductivity agrees with the dc formula computed in \cite{Donos:2019tmo} for this class of holographic models
\begin{equation}\label{sigmadc}
\sigma_{\text{dc}}(\lambda=0)=Z(\phi_h)+\frac{4 \pi \rho^2}{s \ell^2  \tilde{Y}(\phi_h)}\,.
\end{equation}
This constitutes additional evidence that for $\lambda=0$, only momentum relaxes.

 %%%%%%%%%%%%%%%%%
 \subsection{Transverse spectrum at nonzero wavevector}
 \label{tra_sou}
To provide further evidence that the model at $\lambda=0$, $\ell\neq0$ is well described by Wigner crystal hydrodynamics with momentum relaxation only (recall that $\lambda=0$ implies $\Omega=\omega_o=0$), we have computed the transverse QNMs at nonzero wavevector $q$. In the transverse channel, 
Wigner crystal hydrodynamics predicts the existence of two light modes \cite{Delacretaz:2017zxd}, whose dispersion relation is given by: 
 \begin{multline}\label{hydrodispersiontran}
\omega_\pm=\frac1{2}\left[-i\Gamma-iq^2\left(\xi_\perp+\frac{\eta}{\chi_{PP}}\right) \right. \\
\left.\pm \frac{1}{\chi_{PP}} \sqrt{-\Gamma^2\chi_{PP}^2
+2q^2\chi_{PP}\left(2G+\Gamma\xi_\perp\chi_{PP}-\Gamma\eta\right)
-q^4\left(\eta-\xi_\perp\chi_{PP}\right)^2}\right] \ .
 \end{multline}
 
For small $q\ll\Gamma\ll\Lambda$ (denoting by $\Lambda$ the UV cutoff of the hydrodynamic regime), these modes capture transverse phonon diffusion and momentum relaxation, respectively:
 \begin{equation}
 \omega_+=-i\frac{G}{\chi_{PP}\Gamma}q^2-i\xi_\perp q^2+O\left(\frac{q^4}{\Gamma^3}\right)\,,\quad\omega_-=-i\Gamma+i\frac{G}{\chi_{PP}\Gamma}q^2-\frac{\eta}{\chi_{PP}}q^2+O\left(\frac{q^4}{\Gamma^3}\right).
 \end{equation}
 In the limit $\Gamma\ll q\ll\Lambda$, the modes are instead propagating
 \begin{equation}
 \omega_\pm=\pm \sqrt{\frac{G}{\chi_{PP}}}q-\frac{i}2\Gamma+O\left(q\Gamma\right).
 \end{equation}
 Indeed, this regime is probing shorter scales than the scale at which momentum relaxes, so the modes resemble shear sound waves. 
 
 Finally, we can also consider the regime $\Gamma\ll q\simeq\Lambda$, which amounts to sending $q\to+\infty$ in \eqref{hydrodispersiontran}. The modes become purely imaginary once again
 \begin{equation}
 \omega_+=-i\xi_\perp q^2-\frac{iG}{\eta-\chi_{PP}\xi_\perp}+O\left(\frac{q^4}{\Gamma^3}\right)\,,\quad\omega_-=-\frac{i\eta q^2}{\chi_{PP}}+\frac{iG}{\eta-\chi_{PP}\xi_\perp}-i\Gamma+O\left(\frac1{q^2},\frac{\Gamma}{q^2}\right).
 \end{equation}
 Strictly speaking this regime is outside the validity of the hydrodynamic approximation. 
 Yet, as we shall see, the expressions above agree very well with the dispersion relation of the QNMs computed holographically.
 
{In the previous formulae $G$ is the phonon shear modulus, $\eta$ is the shear viscosity and $\xi_{\perp}$ is the diffusivity of the transverse phonon $\varphi_\perp$.} 
These quantities are defined by the hydrodynamic correlators
\begin{eqnarray}
\label{KuboUnrelaxed}
&G^R_{T^{xy}T^{xy}} =G-i\omega\eta\,,\qquad G^R_{\varphi_\perp\varphi_\perp}=
\frac{1}{\chi_{PP}\,\omega^2}-\frac{\xi_\perp}{G}\frac{i}{\omega}\,.
\end{eqnarray}

 \begin{figure}
\begin{center}
\includegraphics[width=0.49\textwidth]{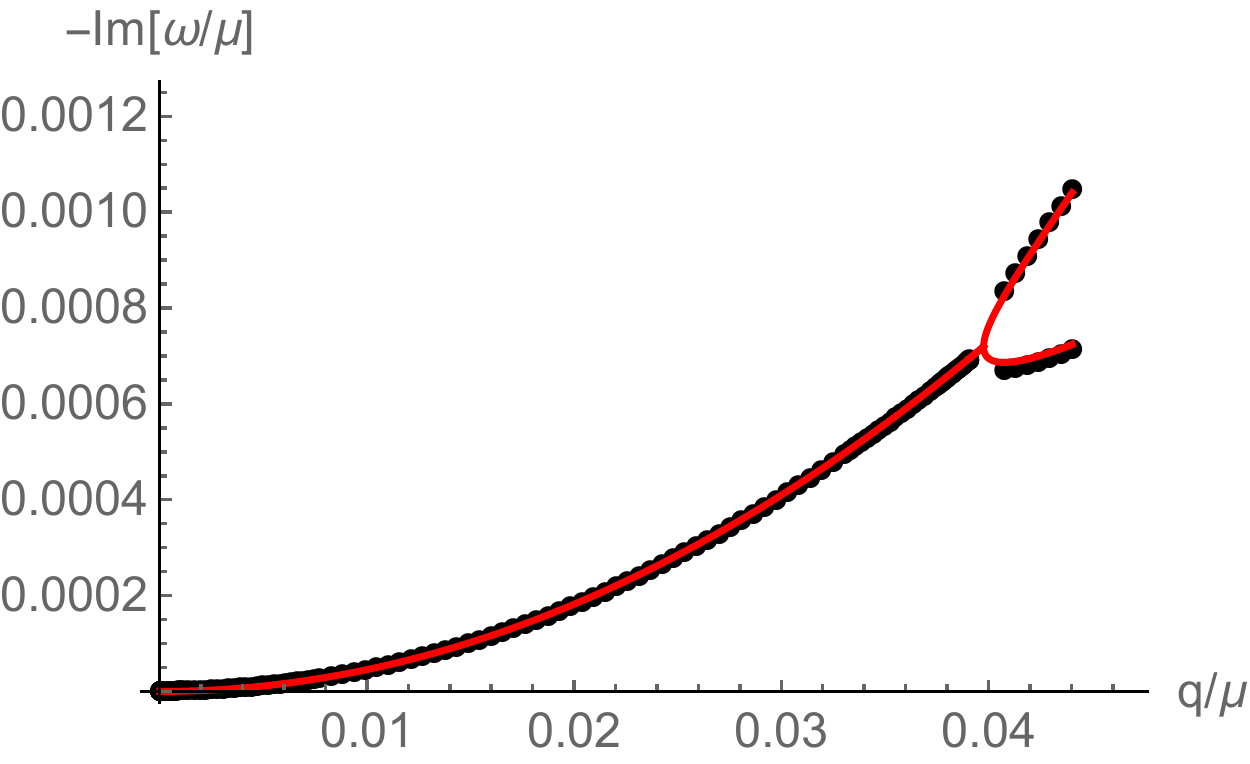}
\includegraphics[width=0.49\textwidth]{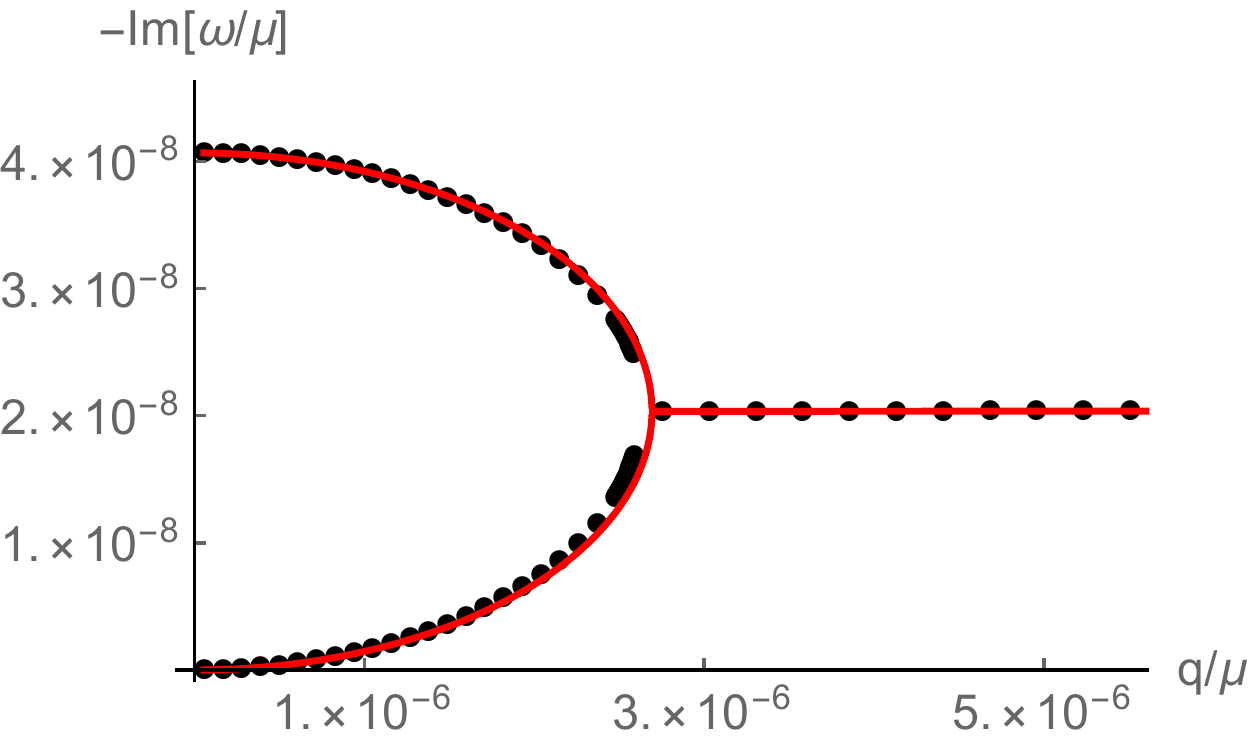}
\includegraphics[width=0.49\textwidth]{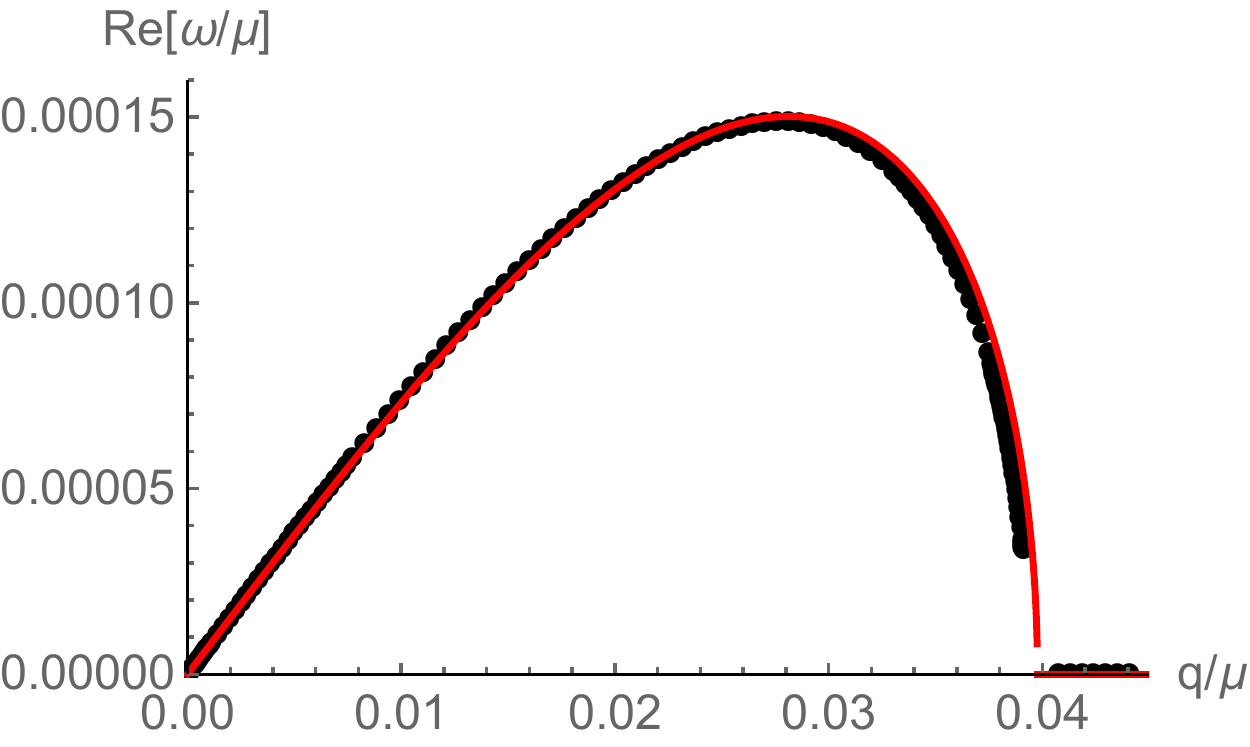}
\includegraphics[width=0.49\textwidth]{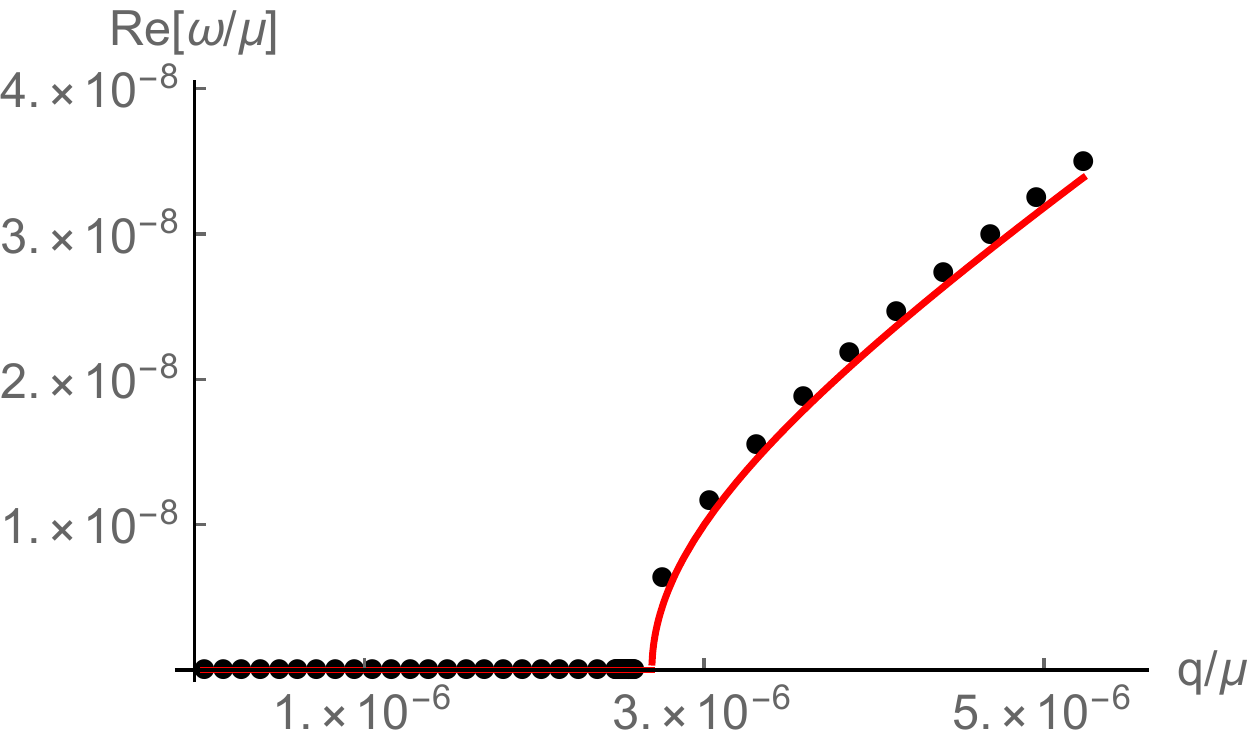}
\end{center}
\caption{Comparison of the hydrodynamic dispersion relation \eqref{hydrodispersiontran} (red lines) to the numerical results (black dots) for large $q$ (left panels) and small $q$ (right panels), at $(\ell/\mu,k/\mu,T/\mu)=(10^{-4},10^{-2},10^{-1})$.}
\label{Re_q}
\end{figure}

In figure \ref{Re_q}, we compare the dispersion relation \eqref{hydrodispersiontran} to the location of the two longest-lived QNMs computed numerically.  In  \eqref{hydrodispersiontran}, we have used for $G$ and $\eta$ the approximate relations \cite{Amoretti:2019cef} (see also \cite{Baggioli:2018bfa,Alberte:2017oqx}):
\begin{equation}\label{approxi}
\eta \simeq \frac{s}{4 \pi} \ , \qquad G\simeq k^2 I_Y \ .
\end{equation}
The prescription for $G$ is further discussed in  subsection \ref{stiffness}.
Due to isotropy,  the diffusivity $\xi_\perp$ is related to $\xi_\parallel$ and is given by the formula, \cite{Amoretti:2019cef}:
\begin{equation}\label{xiper}
\frac{\xi_{\perp}}{G}=\frac{\xi_\parallel}{K+G}=\frac{1}{\chi_{PP}^2}\left[\frac{4\pi \left(sT+\mu\rho\right)^2}{k^2 s Y(\phi_h)}+\mu^2 Z(\phi_h)\right].
\end{equation}
As shown in figure \ref{Re_q} the agreement between the numerics and \eqref{hydrodispersiontran} is excellent. In particular, the three regimes described above can be observed, including the regime $\Gamma\ll q\simeq \Lambda$ where we might have expected the hydrodynamic approximation to fail. This suggests that the true hydrodynamic cutoff in our system is not set by $T$ (observe that the agreement persists to $q\simeq T$) but by some higher-energy scale.\footnote{A more precise characterization of the validity of the hydrodynamic regime would require a detailed study of deviations between the numerical and hydrodynamic dispersion relations, taking into account corrections from terms appearing at higher order in gradients or in the relaxation parameters in the hydrodynamic expansion, and the proximity of other gapped QNMs. See eg \cite{Withers:2018srf,Grozdanov:2019kge}.}

%%%%%%%%%%%%%%%%%
\subsection{The phonon stiffness prescription}
\label{stiffness}

In this section, we highlight a subtlety in the correct determination of the shear phonon stiffness $G$. 
When translations are only broken spontaneously, $G$ can be extracted unambiguously from a Kubo formula involving the shear stress tensor
\begin{equation}
\label{sheareq1}
G=-\lim_{\omega\to0}\textrm{Re}\left[G^R_{T_{xy}T_{xy}}(\omega,q=0)\right].
\end{equation}
This formula directly follows from using the expression in \eqref{KuboUnrelaxed}. 
It states that the phonon shear modulus (stiffness) is equal to the total shear modulus, defined as the response of the system to shear strain.

The shear correlator $G^R_{T_{xy}T_{xy}}$ can be computed holographically by solving the (decoupled) perturbation equation for $h_x^y$.\footnote{This is by now a standard computation. It proceeds exactly along the same lines as described in \cite{Amoretti:2019cef} for our model. See eg \cite{Alberte:2015isw} for earlier versions of this computation.} For small enough $k$, the exact numerical result matches well with a perturbative analytical computation, \cite{Amoretti:2019cef} (see also \cite{Baggioli:2018bfa,Alberte:2017oqx}):
\begin{equation}
\label{sheareq2}
\ell=0,\lambda=0:\quad -\lim_{\omega\to0}\textrm{Re}\left[G^R_{T_{xy}T_{xy}}(\omega,q=0)\right]=k^2 I_{Y}+O(k^4)\,.
\end{equation}
Putting \eqref{sheareq2} together with \eqref{sheareq1}  leads to the identification of the shear phonon stiffness $G$
\begin{equation}
\label{sheareq3}
\ell=0,\lambda=0:\quad G=k^2 I_{Y}+O(k^4)\,.
\end{equation}
We are quoting a formula valid for sufficiently small $k$, but a non-perturbative formula in terms of the $h_{x}^y$ perturbation evaluated at the horizon also exists (see eg \cite{Hartnoll:2016tri}). 
Upon turning weak explicit breaking $\lambda\ll\mu$, the holographic shear correlator computed numerically only receives very small, $O(\lambda)$ corrections and it is straightforward to verify that the approximate formula \eqref{sheareq2} is still valid. 

The explicit breaking scale $\ell$ has a qualitatively different effect.
Upon turning weak explicit breaking $\ell\ll\mu$, the holographic shear Kubo formula above becomes
\begin{equation}\label{approxigellcorr}
-\lim_{\omega\to0}\textrm{Re}\left[G^R_{T_{xy}T_{xy}}(\omega,q=0)\right] = k^2 I_Y+\ell^2 I_{\tilde{Y}} +O(k^4,\ell^4,\ell^2 k^2)\ ,
\end{equation}
where we are still assuming small $k\ll\mu$. $I_Y$ and $I_{\tilde{Y}}$ are given in \eqref{renintegral}. In particular, the right-hand side is nonzero even in absence of any spontaneous breaking $k=0$, as was noted in \cite{Alberte:2015isw}.

As a consequence, identifying the phonon shear modulus $G$ by combining \eqref{sheareq1} and \eqref{approxigellcorr} is no longer correct, and instead the prescription \eqref{sheareq3} should still be used.

For the very small value of $\ell/\mu=10^{-4}$ used in the previous section, we found that the total shear modulus, namely the real part of the correlator $G^R_{T^{xy}T^{xy}}$, is always well approximated by the $k^2$ contribution in \eqref{sheareq2}. 
So the numerical results for the dispersion relation of the transverse modes plotted in figure \ref{Re_q} are still well reproduced by using either \eqref{sheareq2} or the exact numerical result \eqref{sheareq1}.

\begin{figure}
\begin{center}
\includegraphics[width=0.60\textwidth]{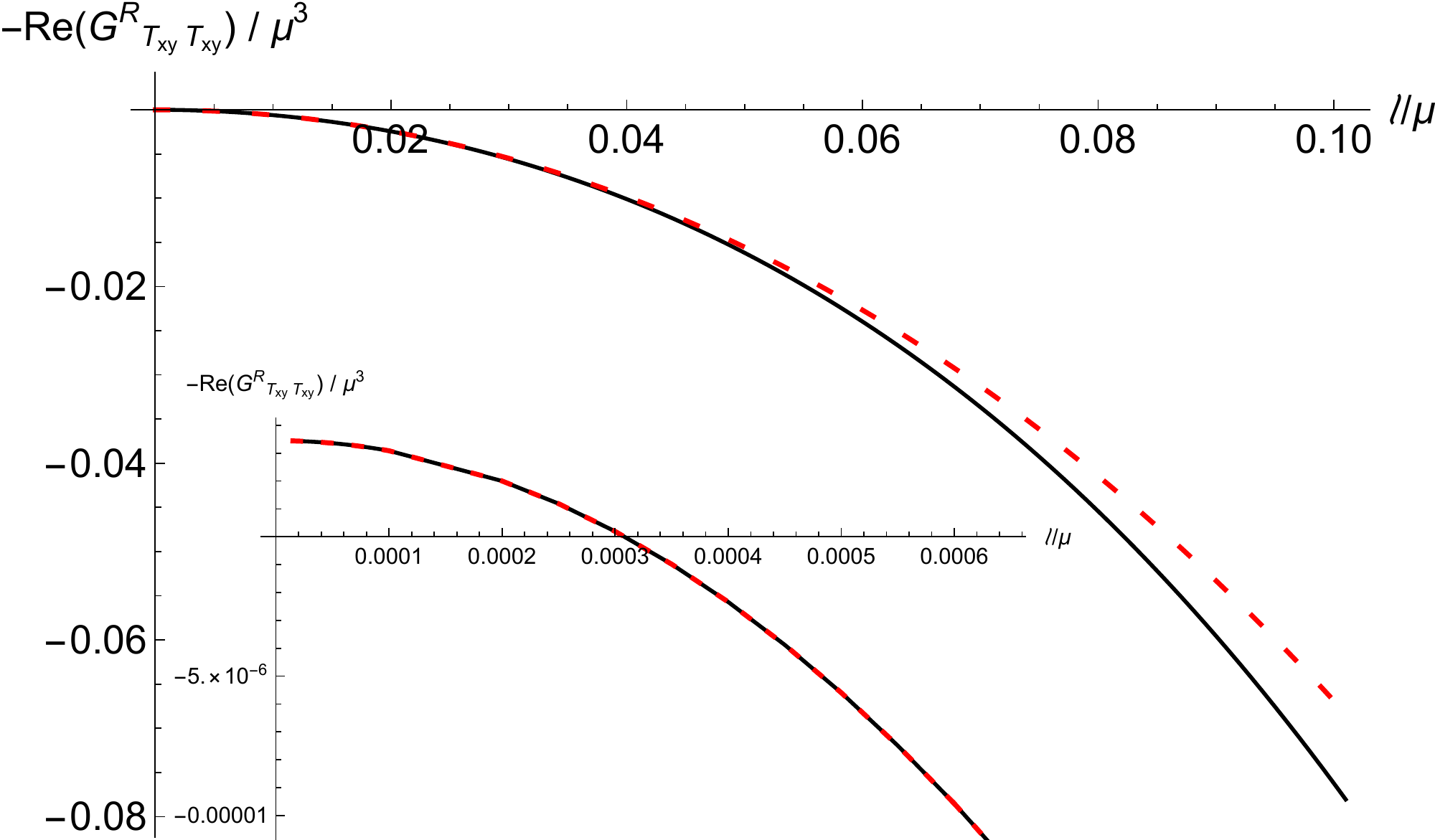}
\end{center}
\caption{Comparison between the total shear modulus $-\lim_{\omega\to0}\textrm{Re}\left[G^R_{T_{xy}T_{xy}}(\omega,q=0)\right]$ computed numerically (black line) and the approximation \eqref{approxigellcorr} (red dashed line) at varying $\ell/\mu$ with $(T/\mu,k/\mu,\lambda/\mu)=(0.1,10^{-3},0)$.}
\label{stiffnessfigure}
\end{figure}

For larger values of $\ell$, the total shear modulus starts to deviate from \eqref{sheareq2} and instead \eqref{approxigellcorr} should be used. 
\eqref{approxigellcorr} agrees very well with the numerics for $\ell$ not too large, see figure \ref{stiffnessfigure}. 
We observe that the first term  in \eqref{approxigellcorr} (proportional to $k^2$) is always positive, 
while the second term (proportional to $\ell^2$) is negative. Accordingly, for large enough $\ell$, the total shear modulus becomes negative. 
As $G$ is positive definite (it is a static susceptibility), the Kubo formula \eqref{sheareq1} is manifestly wrong in this regime. 
This implies that the contribution to the total shear modulus coming from $\ell$ should be removed to identify correctly the phonon shear modulus. 
This is in agreement with the fact that the explicit breaking scale $\ell$ only relaxes momentum, not the phonons, as it does not break the global shift symmetry associated to them.

\begin{figure}
\begin{tabular}{cc}
\includegraphics[width=0.48\textwidth]{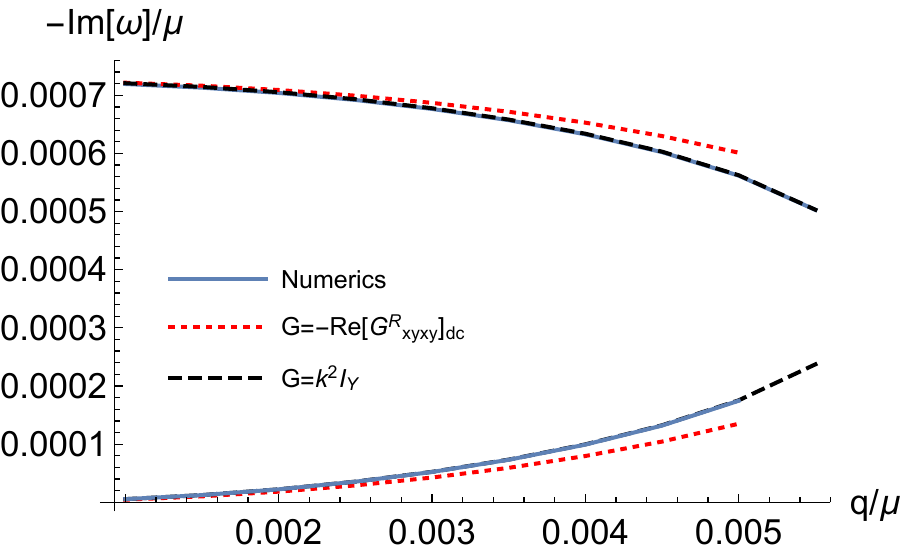}&\includegraphics[width=0.48\textwidth]{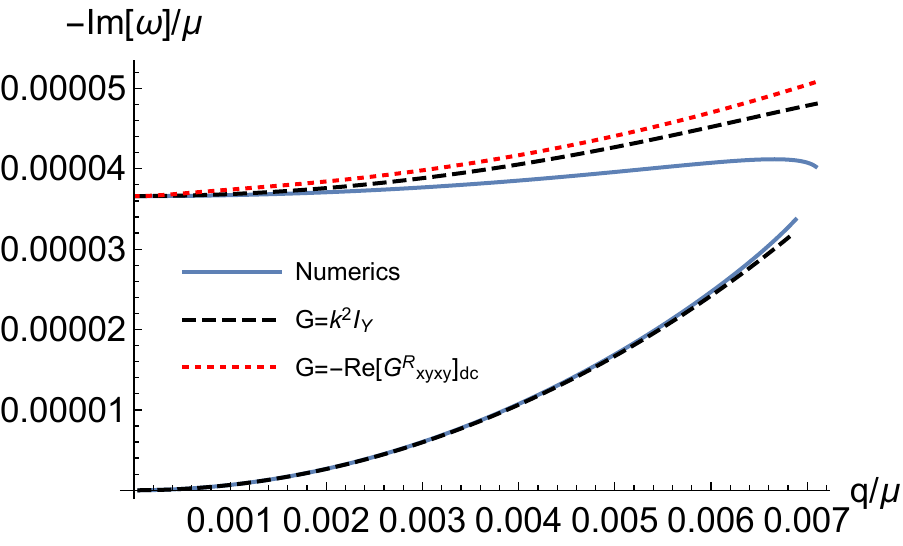}
\end{tabular}
\caption{Comparison between the exact numerical dispersion relation of the two longest-lived transverse QNMs and the hydrodynamic prediction \eqref{hydrodispersiontran} for the dispersion relation, using for the Goldstone shear modulus the correct prescription \eqref{sheareq3} or the wrong one \eqref{sheareq1}. In the right panel, the wrong prescription predicts that the gapless mode is in the upper half plane (not displayed). Left: $\ell/\mu=0.01$, $T/\mu=0.1$, $k/\mu=0.1$. Right: $\ell/\mu=0.003$, $T/\mu=0.1$, $k/\mu=0.001$}
\label{HydrodispcompG}
\end{figure}

In figure \ref{HydrodispcompG} we compare the dispersion relation of the transverse QNMs for two sets of values $(\ell/\mu,T/\mu,k/\mu)=(0.01,0.1,0.1)$  and $(\ell/\mu,T/\mu,k/\mu)=(0.003,0.1,0.001)$ to the hydrodynamic prediction \eqref{hydrodispersiontran} using for the phonon shear modulus either the correct prescription \eqref{sheareq3} or the wrong one \eqref{sheareq1}. For the first case, the total shear modulus is positive, but the prescription \eqref{sheareq3} gives significantly better results. For the second case, the total shear modulus is now negative. The prescription \eqref{sheareq1} predicts that the gapless mode is in the upper half plane (not displayed) and does not agree very well with the gapped mode. Instead, the prescription \eqref{sheareq3} gives a very good account of the numerical result for the gapless mode, and agrees better with the gapped mode.

Our observation that the explicit breaking scale $\ell$ contributes to the total shear modulus with the opposite sign compared to the spontaneous scale $k$ is consistent with previous literature \cite{Andrade:2013gsa,Alberte:2015isw,Alberte:2016xja,Burikham:2016roo,Andrade:2017ghg,Alberte:2017cch,Alberte:2017oqx,Baggioli:2018bfa,Amoretti:2018tzw,Andrade:2018}.

%%%%%%%%%%%%%%%%%%%%%%%%
\section{Gapped phonons}
\label{section:ellandlambdanonzero}

\subsection{Spectrum at zero wavevector}
\begin{figure}
\begin{center}
\includegraphics[width=0.49\textwidth]{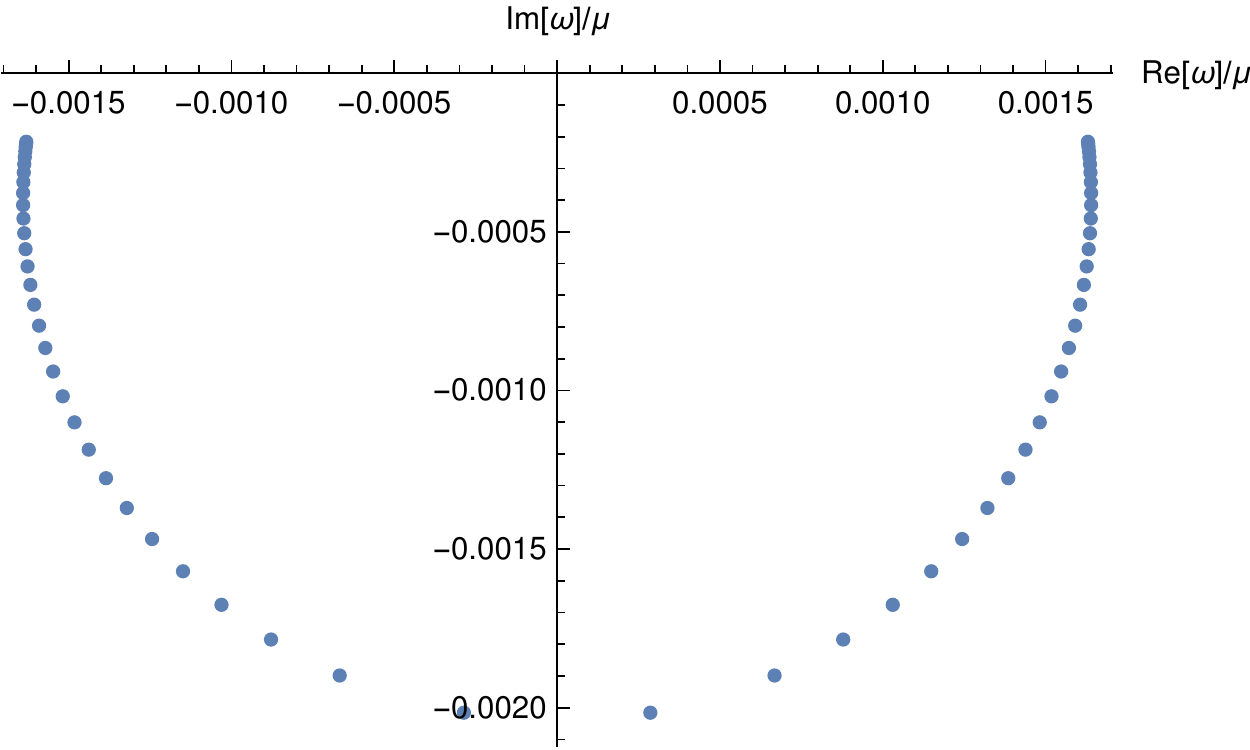}
\end{center}
\caption{ $\lambda/\mu=0.0002$, $k/\mu=0.1$, $T/\mu=0.1$ and  $0.001\leq\ell/\mu\leq0.03$. As $\ell/\mu$ increases, the complex QNMs move down the lower half plane and approach the imaginary axis, where they eventually collide.}
\label{fig:QNMselllbnonzero}
\end{figure}

Once a small source $\lambda$ for the field $\phi$ is switched on, we expect the phonons in our holographic system to become massive and damped. Indeed, we plot the lightest QNMs of the system in figure \ref{fig:QNMselllbnonzero} and observe that now two gapped, complex QNMs are present close to the real axis at small $\ell/\mu$. Increasing $\ell/\mu$, the QNMs move down in the lower half plane, getting closer to the imaginary axis, where a collision eventually happens at large enough $\ell/\mu$ (not displayed).
The dc conductivity can be computed holographically \cite{Donos:2019tmo}:
\begin{equation}
\label{dccondelllambdanonzero}
\sigma_{\text{dc}}(\lambda\neq0,\ell\neq0)=Z(\phi_h)+\frac{4 \pi \rho^2}{s( \ell^2  \tilde{Y}(\phi_h)+k^2 Y(\phi_h))}\,.
\end{equation}
The ac conductivity predicted by Wigner crystal hydrodynamics assumes the form:
\begin{equation}\label{conduacfinitelambda}
\sigma(\omega)=\sigma_o+\frac{\gamma_1^2 \chi_{PP}^2 \omega_o^2 (\Gamma -i \omega )+2 \gamma_1 \rho  \chi_{PP}  \omega_o^2+\rho ^2 (i \omega - \Omega)}
{\chi_{PP}   \left((\Gamma-i \omega)(\Omega-i \omega)+\omega_o^2\right)} \ .
\end{equation}
It predicts two poles located at (complex) frequencies given by \eqref{wchydrogapped}.
\begin{figure}
\begin{center}
\includegraphics[width=0.49\textwidth]{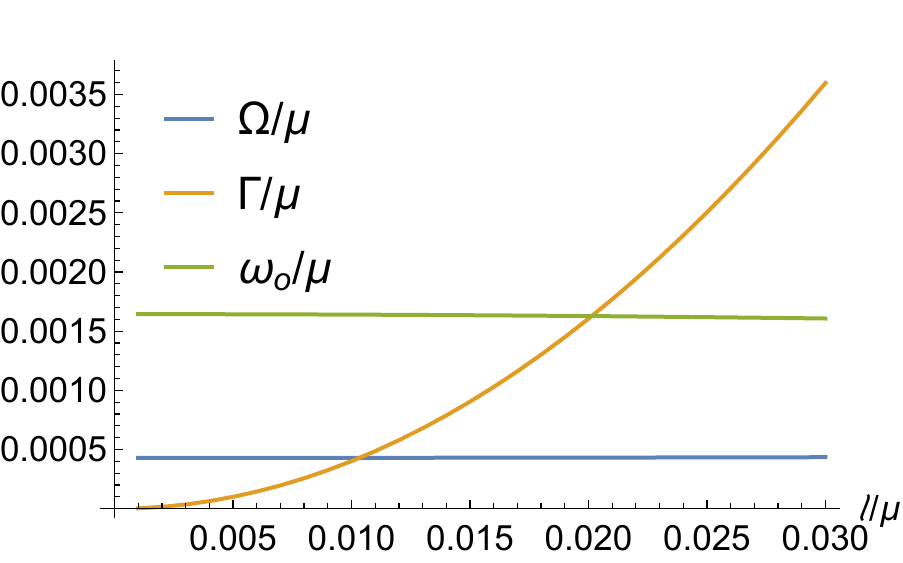}\includegraphics[width=0.49\textwidth]{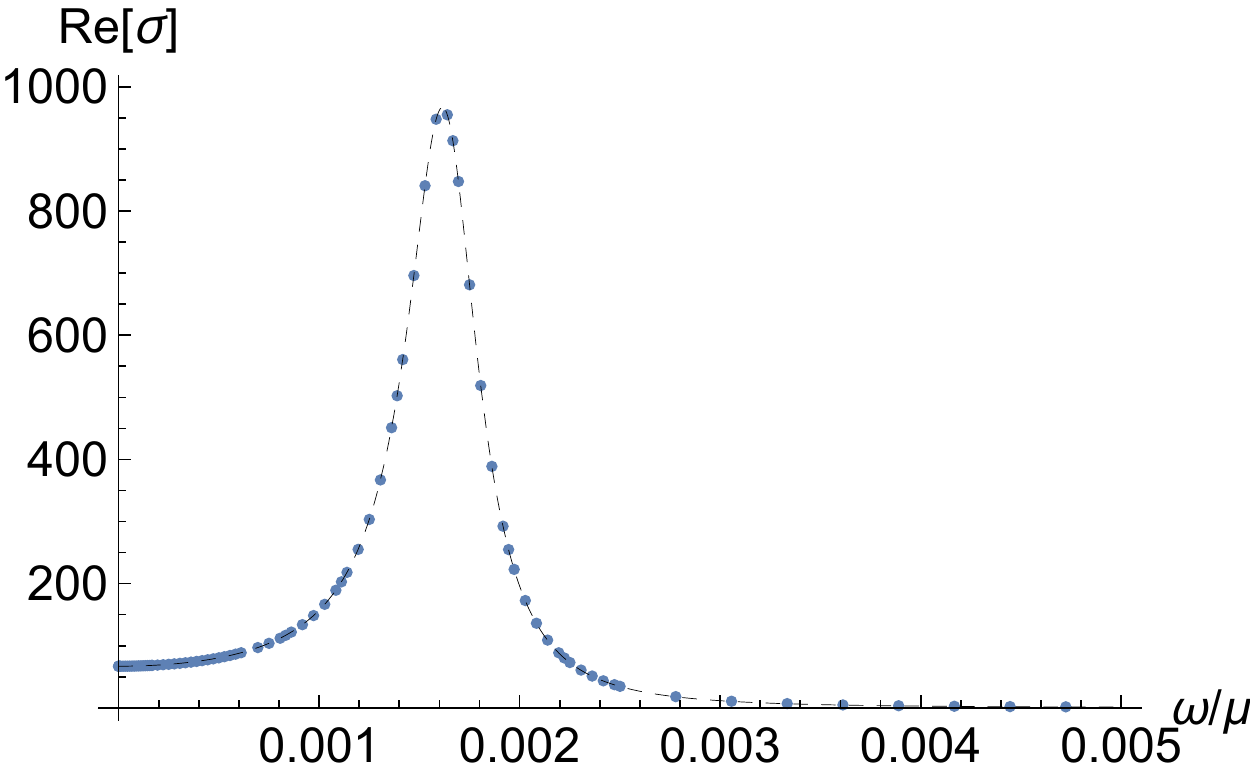}
\end{center}
\caption{$(\lambda/\mu,k/\mu,T/\mu)=(0.0002,0.1,0.1)$. Left: $\ell$ dependence of the relaxation parameters extracted from the two lightest QNMs and the dc conductivity. Right: Real part of the ac electric conductivity computed numerically (blue dots) together with the hydrodynamic prediction \eqref{conduacfinitelambda} (black dashed line) for  $\ell/\mu=0.001$.}
\label{condufinitelambda}
\end{figure}
We have evaluated $\chi_{PP}$ and $\sigma_o$ using \eqref{chipp} and \eqref{sigmainc}, while $\gamma_1$ is given by \cite{Amoretti:2019cef}: 
\begin{equation}
\gamma_1=-\frac{1}{\chi_{PP}^2}\left[\mu (sT+k^2 I_Y) Z(\phi_h)+\frac{4 \pi I_Y \rho (sT+\mu \rho)}{s Y(\phi_h)} \right] \ .
\end{equation}
We identify $\omega_o$, $\Gamma$ and $\Omega$ by matching the dc conductivity and pole locations computed holographically and within hydrodynamics, which fixes all the parameters in \eqref{conduacfinitelambda}. The relaxation parameters are displayed in the left panel  of figure \ref{condufinitelambda}.  
The numerics and the hydrodynamic prediction agree very well, see the right panel of figure \ref{condufinitelambda}. 
We also display the dependence of the relaxation  parameters on the scale $\ell$. As in the case $\lambda=0$, $\Gamma$ shows a clear quadratic dependence. 
On the other hand, $\Omega$ and $\omega_o$ depend very weakly on $\ell$. From \eqref{wchydrogapped}, it is clear that if $\Gamma$ becomes large enough compared to $\omega_o$, the complex poles collide and become purely imaginary, which is what we observe in figure \ref{fig:QNMselllbnonzero}.

\begin{figure}
\begin{center}
\includegraphics[width=0.49\textwidth]{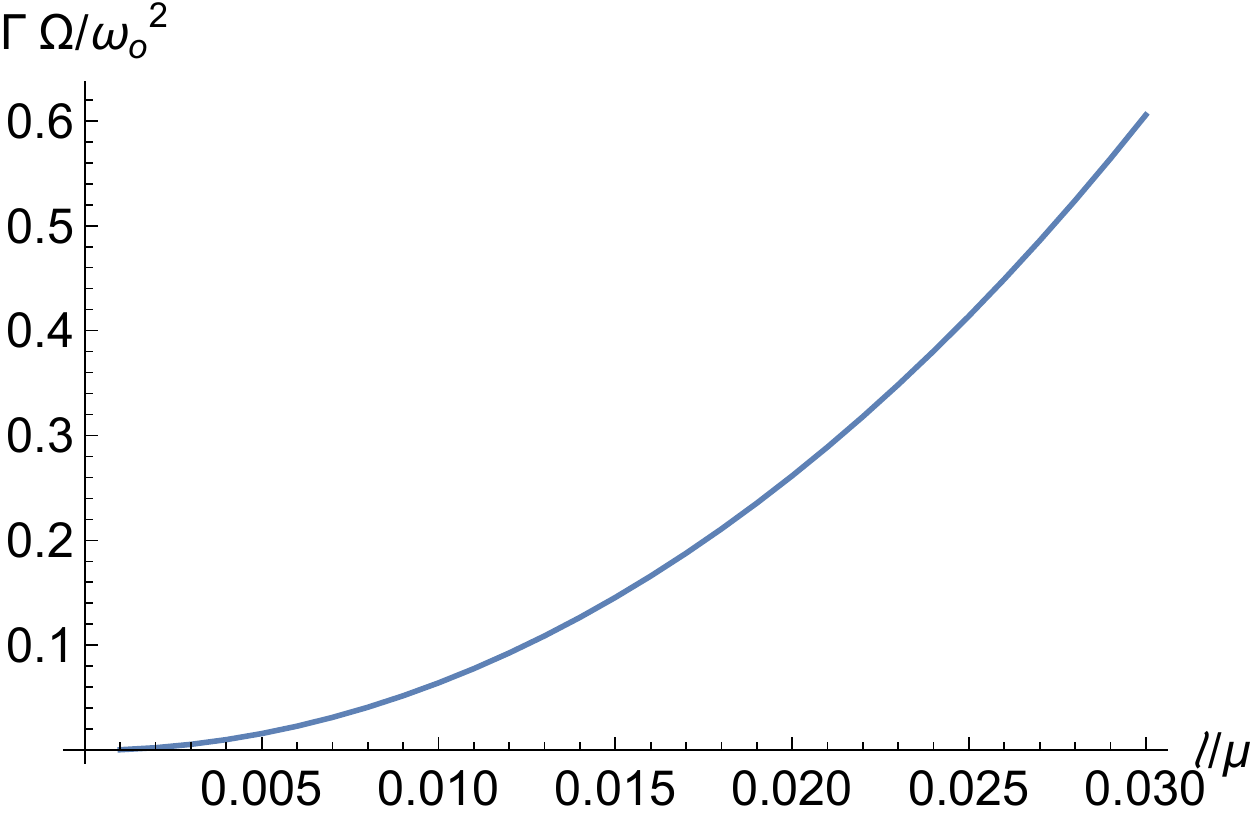}
\includegraphics[width=0.49\textwidth]{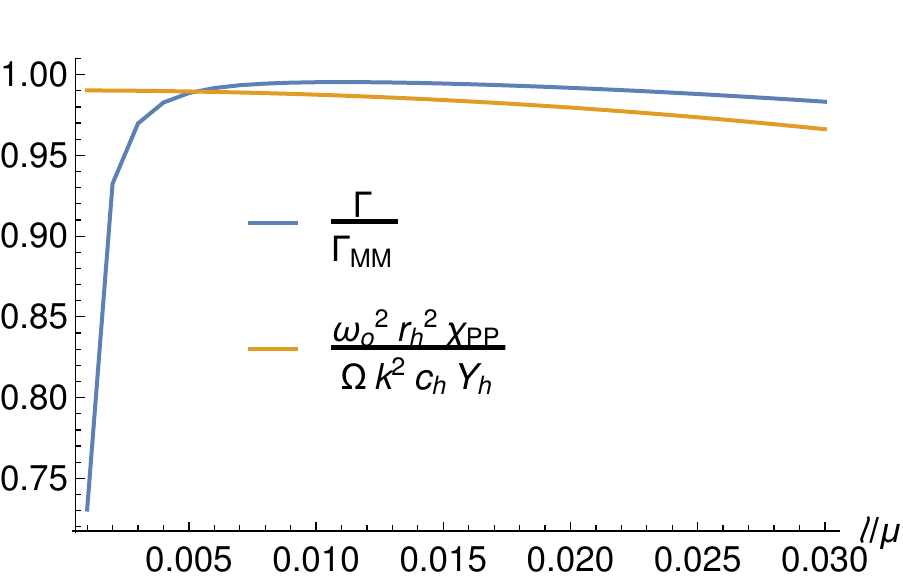}
\end{center}
\caption{$(\lambda/\mu,k/\mu,T/\mu)=(2\times10^{-4},0.1,0.1)$. Left: $\ell$ dependence of the ratio $\Gamma/(\omega_o^2/\Omega)$. Right: Comparison between $\Gamma$, $\omega_o^2/\Omega$ and their approximate memory matrix expressions.}
\label{fig:comprelparam}
\end{figure}

Since $\ell \neq 0$, in this model $\Gamma$ is comparable to $\Omega$ and $\omega_o$ (see left panel of figure \ref{fig:comprelparam}), in contrast to our previous work \cite{Amoretti:2018tzw}, where $\ell=0$ and we found $\Gamma\ll\omega_o,\Omega$. This lets us verify the validity of Wigner crystal hydrodynamics in the presence of a non-negligible source of momentum relaxation, which can be tuned independently from the leading source of phonon damping and pinning, $\lambda$.

We saw in the previous section that when $\lambda=0$, $\Gamma$ was very well approximated by the memory matrix prediction $\Gamma_\text{MM}$ in \eqref{GMM}. In \cite{Amoretti:2018tzw}, setting $\ell=0$, we had shown that $\omega_o^2/\Omega$ was very well approximated by another memory matrix prediction
\begin{equation}
\label{MMomega0ovOmega}
\frac{\omega_o^2}{\Omega}=\frac{k^2 c_h Y(\phi_h)}{r_h^2\chi_{PP}}\,.
\end{equation}
This captures the phonon contribution to momentum relaxation.  In \cite{Amoretti:2018tzw}, with $\ell=0$, $\Gamma$ was negligible and so momentum was solely relaxed through phonon damping. As can be seen in the left panel of figure \ref{fig:comprelparam}, this is no longer the case with $\ell\neq0$.

In the right panel of figure \ref{fig:comprelparam}, we compare the predictions \eqref{GMM} and \eqref{MMomega0ovOmega} to the actual numerical results, at nonzero $\ell$ and $\lambda$. 
While \eqref{MMomega0ovOmega} holds reasonably well at small $\ell$, the agreement steadily gets worse as $\ell$ is increased. 
On the other hand, while $\Gamma$ agrees reasonably well with $\Gamma_\text{MM}$ at larger values of $\ell$ 
(but still low enough that momentum can be considered to be relaxing slowly), there is a sharp disagreement at very low values of $\ell$, 
as the numerical value of $\Gamma$ becomes very small. This is in agreement with our results of \cite{Amoretti:2018tzw} at $\ell=0$, 
where we always observed $\Gamma\ll\omega_o^2/\Omega$.

Having two different sources of explicit translation symmetry breaking, namely $\lambda$ (the source for $\phi$) and $\ell$, 
it is interesting to check the validity of the universal relaxation relation we found in \cite{Amoretti:2018tzw}, 
which relates the ratio of phonon damping rate $\Omega$ and the phonon mass $m$ to a combination of thermodynamic and hydrodynamic data:
\begin{equation}\label{unirel}
\Omega\simeq Gm^2\frac{\xi_\parallel}{K+G}=G m^2 \Xi= \chi_{PP} \omega_o^2 \Xi  \ .
\end{equation}
 In the last equality, we have used the relation between the pinning frequency and the phonon mass $\omega_o^2=m^2 G/\chi_{PP}$. 
 
 In \cite{Amoretti:2018tzw}, we observed (in a slightly different model) that the relation \eqref{unirel} is spoiled as $\lambda$ is increased and that the parameters $\Omega$ and $m$ take values well-approximated by their $k=0$ limit. We explained this by setting $k=0$ and observing that $\Omega$ corresponds to the gap of the longest-lived purely imaginary QNM, which is interpreted as the gapped Goldstone of the pseudo-spontaneously broken global U(1) symmetry of our model (see also \cite{Donos:2019txg}). By increasing $\lambda$, this QNM collides with another QNM coming up from deeper in the lower half plane, after which both QNMs acquire a real part, signalling the exit of the pseudo-spontaneous regime. At this point, the relation \eqref{unirel} fails.
 
\begin{figure}
\begin{center}
\includegraphics[width=0.49\textwidth]{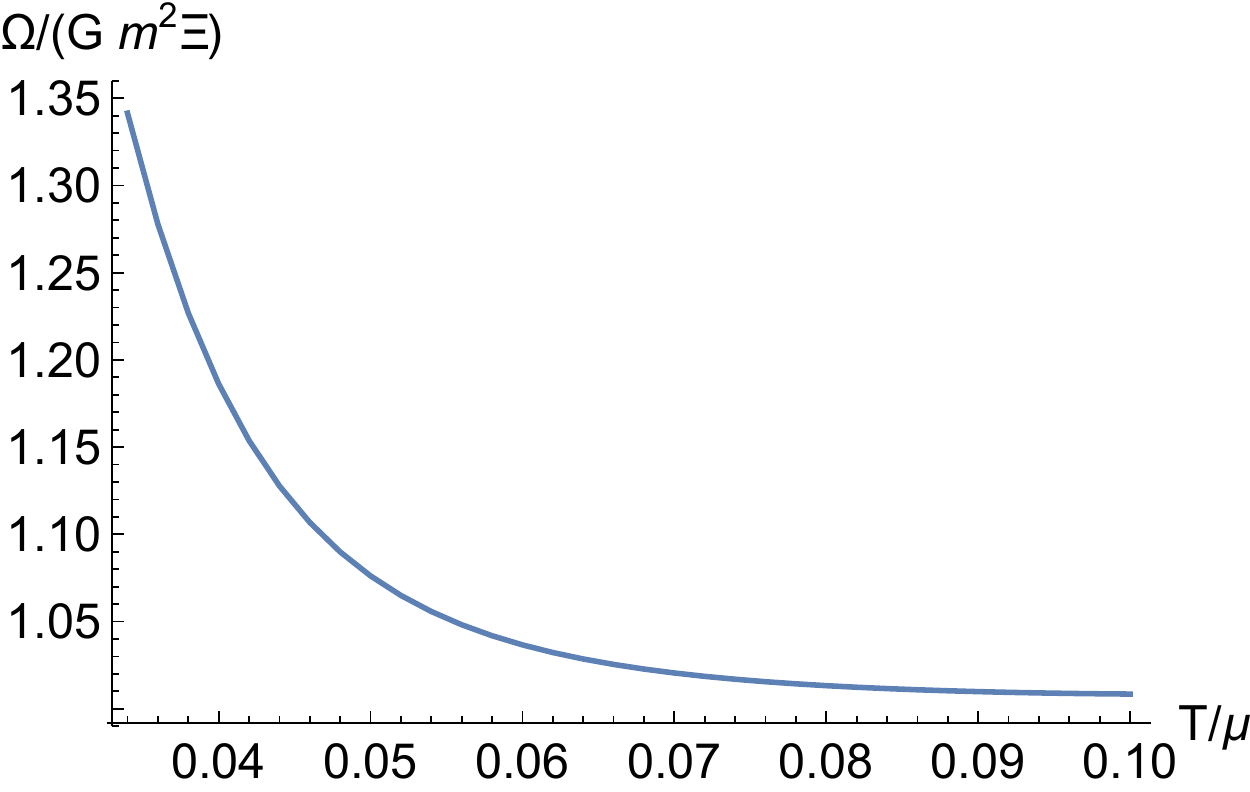}
\includegraphics[width=0.49\textwidth]{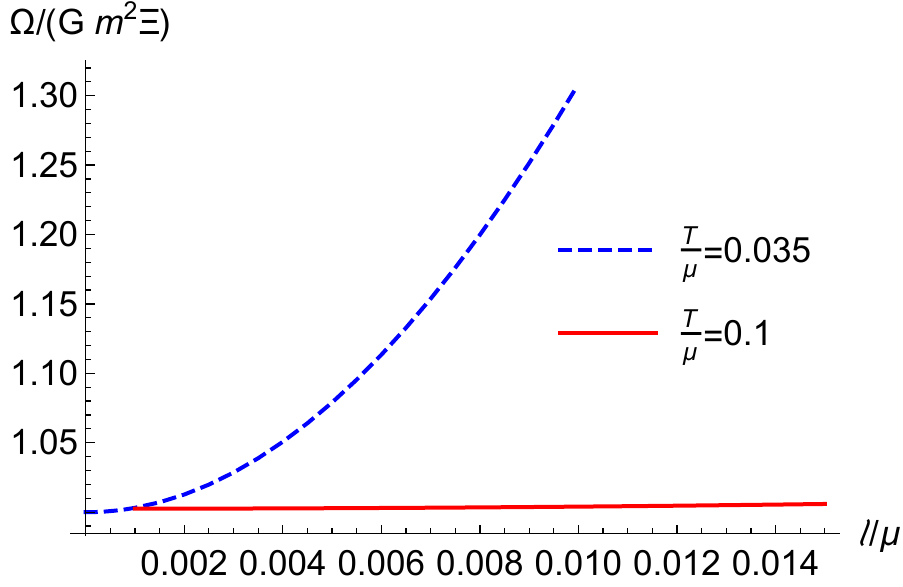}
\end{center}
\caption{Left: $T/\mu$ dependence of the ratio $\Omega/(m^2 \Xi)$ for $(\ell/\mu,k/\mu,\lambda/\mu)=(0.01,0.01,10^{-5})$. 
Right: $\ell/\mu$ dependence of the ratio $\Omega/(m^2 \Xi)$ for $(\lambda/\mu,T/\mu,k/\mu)=(2\times 10^{-4},0.1,0.1)$ (red, solid) and $(\lambda/\mu,T/\mu,k/\mu)=(10^{-5},0.035,0.01)$ (blue, dashed).}
\label{comparizonmxifig}
\end{figure}

In the present work, we have tested the validity of \eqref{unirel} when $T$ and $\ell$ are varied. 
The results are shown in figure \ref{comparizonmxifig}. At high $T$, \eqref{unirel} is almost insensitive to increasing $\ell$. As $T$ is decreased at fixed $\ell\neq0$ and $\lambda\neq0$, the relation \eqref{unirel} ceases to hold. At low $T$, \eqref{unirel} is recovered as $\ell$ is decreased. 

The following interpretation of these results can be offered. At high $T$ and in absence of any spontaneous breaking of translations, $\ell$ is a marginally relevant deformation of the UV CFT and sources weak momentum relaxation at a small rate $\Gamma\ll T,\mu$, which is well-approximated by the memory matrix prediction \cite{Hartnoll:2012rj,Davison:2014lua}. As $T$ is lowered, a coherent-to-incoherent crossover is expected, sometimes accompanied by a pole collision, at which point all notion of almost-conserved momentum is lost. Our case here is more complicated, since momentum mixes with the Goldstone mode when $k\neq0$, but a similar logic applies: at lower temperatures, higher values of $\ell$ lead to a faster relaxation of both momentum and the Goldstone, and consequently to a failure of the relation \eqref{unirel}, which only holds in the limit of slow relaxation. Indeed we have checked that at low temperatures $T/\mu=0.035$, the relaxation parameters (ie $\Gamma$, $\omega_o^2/\Omega$ or their sum) are not well-approximated by any memory matrix expression (ie the right-hand side of \eqref{GMM}, of \eqref{MMomega0ovOmega} or their sum).

\subsection{Transverse spectrum at nonzero wavevector}

In the presence of momentum relaxation, phonon pinning and damping, Wigner crystal hydrodynamics predicts the following dispersion relation \eqref{hydrodispersiontran} for the transverse modes:
\begin{multline}\label{hydrodiffusivefl}
\omega_{\pm}=-\frac{i}{2}\left(\Gamma+\Omega+q^2\left(\xi_{\perp}+\frac{\eta}{\chi_{PP}}\right)\right)\pm\\
\frac{1}{2 \chi_{PP}}\sqrt{\chi_{PP}^2\left(4 \omega_o^2-(\Gamma-\Omega)^2\right)+\chi_{PP} q^2(4G+2(\xi_{\perp}\chi_{PP}-\eta)(\Gamma-\Omega))-q^4(\eta-\xi_{\perp} \chi_{PP})^2} \ .
\end{multline}
\begin{figure}[H]
\begin{center}
\includegraphics[width=0.49\textwidth]{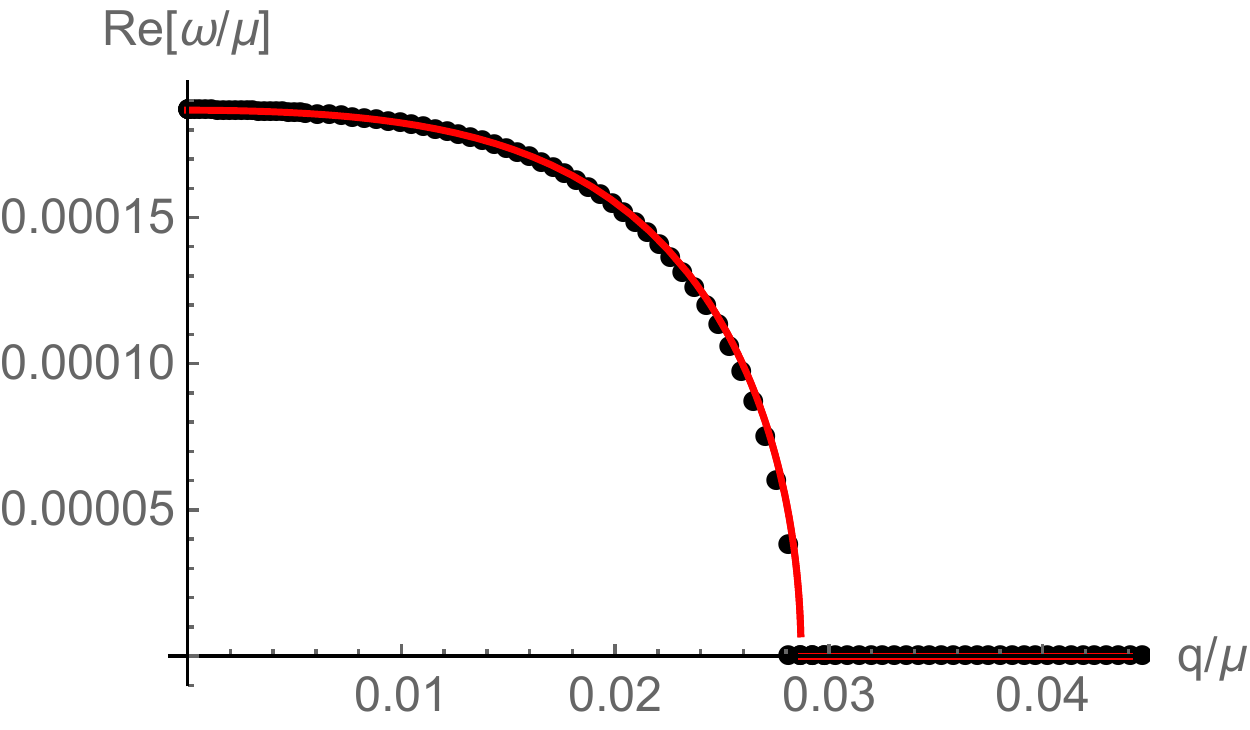}
\includegraphics[width=0.49\textwidth]{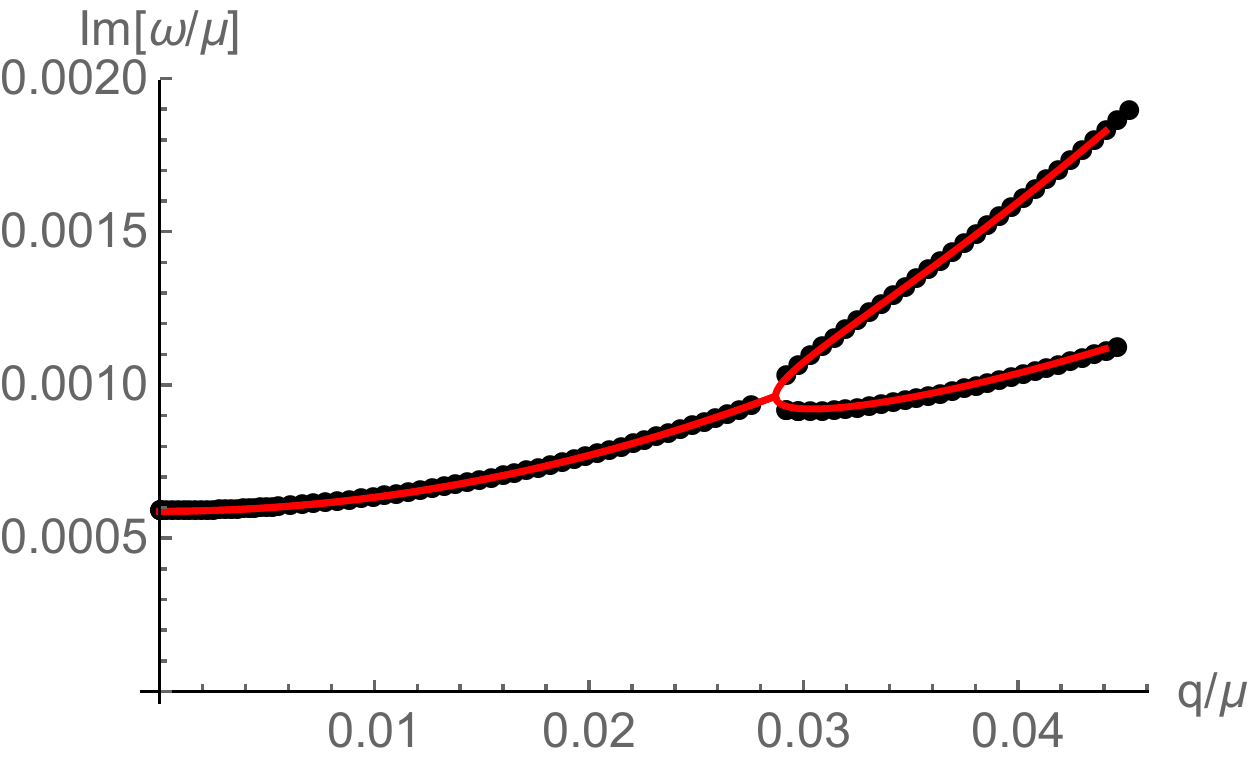}
\end{center}
\caption{ Real (left) and imaginary (right) part of the QNMs computed numerically (black dots) together with the hydrodynamic prediction \eqref{hydrodiffusivefl} (red line) as a function of the momentum $q$ for $(\lambda/\mu,k/\mu,\ell/\mu,T/\mu)=(4\times10^{-4},0.01,0.01,0.1)$.}
\label{qnmtransversefl}
\end{figure}
We have checked the validity of \eqref{hydrodiffusivefl} against the dispersion relation of the QNMs obtained numerically for our model, 
using the same procedure as before to extract the parameters $\omega_o$, $\Gamma$ and $\Omega$ and \eqref{approxi}, \eqref{xiper} for the values of $G$, $\eta$ and $\xi_\perp$. 
The result is displayed in figure \ref{qnmtransversefl} and shows an excellent agreement of the hydrodynamic prediction with the numerical computation. 
For the parameter values we have chosen, due to the presence of phonon pinning and damping, the modes remain complex in the limit of zero wavevector with a gap given by \eqref{wchydrogapped}.

\section{Outlook\label{sec:outlook}}

In this work, we have extended our previous results \cite{Amoretti:2018tzw,Amoretti:2019cef} on {holographic phases which break translations spontaneously} 
by including a non-negligible source of momentum relaxation. 
Indeed, {although \cite{Amoretti:2018tzw} already considered the presence of a small explicit translation-breaking source, momentum relaxation was found to be negligible compared to phonon damping and pinning.}
{By comparing the ac conductivity and the spectrum of transverse QNMs, 
we have verified that the low energy dynamics of the system is described by Wigner crystal hydrodynamics with excellent accuracy.}

In passing, we have discussed a subtlety in the identification of the phonon stiffness from the shear Kubo formula. 
In general, the shear Kubo formula receives an extra contribution from the source of momentum relaxation, which needs to be subtracted. 
Failing to do so leads to inconsistencies upon matching to Wigner crystal hydrodynamics.

Finally, we have shown that the universal relation uncovered in \cite{Amoretti:2018tzw} between the phonon damping rate, 
mass and diffusivities continues to hold in the presence of this new source of explicit translation symmetry breaking, provided it is weak enough.

In the future, it would be interesting to study the longitudinal sector of these theories and their match to Wigner crystal hydrodynamics. Some disagreement in holographic massive gravity models has been reported \cite{Ammon:2019apj}. {On the other hand, \cite{Armas:2019sbe} recently carried out a careful analysis of the dispersion relation of the modes with a nonzero background strain, and find new contributions in the longitudinal sector, which may resolve this discrepancy.}\footnote{Based on private communication with the authors of \cite{Ammon:2019apj} and \cite{Armas:2019sbe}.}

Finally, the universal relation \eqref{unirel} remains to be tested more generally, in other holographic models of spontaneous translation symmetry breaking, homogeneous or not \cite{Andrade:2017cnc,Andrade:2017ghg,Andrade:2018}, {in particular for thermodynamically stable phases}, or directly in field theory.

\begin{acknowledgments}
BG would like to thank Akash Jain for discussions.
BG would like to thank MPI-PKS for warm hospitality during the final stages of this work. DA would like to thank CPHT at Ecole Polytechnique for warm hospitality during the final stages of this work.	
D.A. is supported by the `Atracci\'on del Talento' programme (Comunidad de Madrid) under grant 2017-T1/TIC-5258 and by MCIU/AEI/FEDER, UE, through the grants SEV-2016-0597, FPA2015-65480-P and PGC2018-095976-B-C21.
B.G. is supported by the European Research Council (ERC) under the European Union’s Horizon 2020 research and innovation programme (grant agreement No 758759).
The work of DM has been funded by the Spanish grants FPA2014-52218-P and
FPA2017-84436-P by Xunta de Galicia (GRC2013-024), by FEDER 
and by the Mar\'ia de Maeztu Unit of Excellence MDM-2016-0692. D.A. and D.M. thank the FRont Of pro-Galician Scientists for unconditional support. 

\end{acknowledgments}	

\bibliographystyle{JHEP}
\bibliography{MTSB}

\providecommand{\href}[2]{#2}\begingroup\raggedright\begin{thebibliography}{10}

\bibitem{Amoretti:2018tzw}
A.~Amoretti, D.~Areán, B.~Goutéraux and D.~Musso, \emph{{Universal relaxation
  in a holographic metallic density wave phase}},
  \href{https://doi.org/10.1103/PhysRevLett.123.211602}{\emph{Phys. Rev. Lett.}
  {\bfseries 123} (2019) 211602}
  [\href{https://arxiv.org/abs/1812.08118}{{\ttfamily 1812.08118}}].

\bibitem{chaikin_lubensky_1995}
P.~M. Chaikin and T.~C. Lubensky, \emph{Principles of Condensed Matter
  Physics}. Cambridge University Press, 1995,
  \href{https://doi.org/10.1017/CBO9780511813467}{10.1017/CBO9780511813467}.

\bibitem{Hartnoll:2016apf}
S.~A. Hartnoll, A.~Lucas and S.~Sachdev, \emph{{Holographic quantum matter}},
  \href{https://arxiv.org/abs/1612.07324}{{\ttfamily 1612.07324}}.

\bibitem{Horowitz:2012ky}
G.~T. Horowitz, J.~E. Santos and D.~Tong, \emph{{Optical Conductivity with
  Holographic Lattices}},
  \href{https://doi.org/10.1007/JHEP07(2012)168}{\emph{JHEP} {\bfseries 07}
  (2012) 168} [\href{https://arxiv.org/abs/1204.0519}{{\ttfamily 1204.0519}}].

\bibitem{Chesler:2013qla}
P.~Chesler, A.~Lucas and S.~Sachdev, \emph{{Conformal field theories in a
  periodic potential: results from holography and field theory}},
  \href{https://doi.org/10.1103/PhysRevD.89.026005}{\emph{Phys. Rev.}
  {\bfseries D89} (2014) 026005}
  [\href{https://arxiv.org/abs/1308.0329}{{\ttfamily 1308.0329}}].

\bibitem{Hartnoll:2014cua}
S.~A. Hartnoll and J.~E. Santos, \emph{{Disordered horizons: Holography of
  randomly disordered fixed points}},
  \href{https://doi.org/10.1103/PhysRevLett.112.231601}{\emph{Phys. Rev. Lett.}
  {\bfseries 112} (2014) 231601}
  [\href{https://arxiv.org/abs/1402.0872}{{\ttfamily 1402.0872}}].

\bibitem{Donos:2014yya}
A.~Donos and J.~P. Gauntlett, \emph{{The thermoelectric properties of
  inhomogeneous holographic lattices}},
  \href{https://doi.org/10.1007/JHEP01(2015)035}{\emph{JHEP} {\bfseries 01}
  (2015) 035} [\href{https://arxiv.org/abs/1409.6875}{{\ttfamily 1409.6875}}].

\bibitem{Donos:2017ihe}
A.~Donos, J.~P. Gauntlett and V.~Ziogas, \emph{{Diffusion for Holographic
  Lattices}}, \href{https://doi.org/10.1007/JHEP03(2018)056}{\emph{JHEP}
  {\bfseries 03} (2018) 056}
  [\href{https://arxiv.org/abs/1710.04221}{{\ttfamily 1710.04221}}].

\bibitem{Donos:2012js}
A.~Donos and S.~A. Hartnoll, \emph{{Interaction-driven localization in
  holography}}, \href{https://doi.org/10.1038/nphys2701}{\emph{Nature Phys.}
  {\bfseries 9} (2013) 649} [\href{https://arxiv.org/abs/1212.2998}{{\ttfamily
  1212.2998}}].

\bibitem{Vegh:2013sk}
D.~Vegh, \emph{{Holography without translational symmetry}},
  \href{https://arxiv.org/abs/1301.0537}{{\ttfamily 1301.0537}}.

\bibitem{Donos:2013eha}
A.~Donos and J.~P. Gauntlett, \emph{{Holographic Q-lattices}},
  \href{https://doi.org/10.1007/JHEP04(2014)040}{\emph{JHEP} {\bfseries 04}
  (2014) 040} [\href{https://arxiv.org/abs/1311.3292}{{\ttfamily 1311.3292}}].

\bibitem{Andrade:2013gsa}
T.~Andrade and B.~Withers, \emph{{A simple holographic model of momentum
  relaxation}}, \href{https://doi.org/10.1007/JHEP05(2014)101}{\emph{JHEP}
  {\bfseries 05} (2014) 101} [\href{https://arxiv.org/abs/1311.5157}{{\ttfamily
  1311.5157}}].

\bibitem{Gouteraux:2014hca}
B.~Gout\'eraux, \emph{{Charge transport in holography with momentum
  dissipation}}, \href{https://doi.org/10.1007/JHEP04(2014)181}{\emph{JHEP}
  {\bfseries 04} (2014) 181} [\href{https://arxiv.org/abs/1401.5436}{{\ttfamily
  1401.5436}}].

\bibitem{Baggioli:2014roa}
M.~Baggioli and O.~Pujolas, \emph{{Electron-Phonon Interactions,
  Metal-Insulator Transitions, and Holographic Massive Gravity}},
  \href{https://doi.org/10.1103/PhysRevLett.114.251602}{\emph{Phys. Rev. Lett.}
  {\bfseries 114} (2015) 251602}
  [\href{https://arxiv.org/abs/1411.1003}{{\ttfamily 1411.1003}}].

\bibitem{Donos:2014oha}
A.~Donos, B.~Gout\'eraux and E.~Kiritsis, \emph{{Holographic Metals and
  Insulators with Helical Symmetry}},
  \href{https://doi.org/10.1007/JHEP09(2014)038}{\emph{JHEP} {\bfseries 09}
  (2014) 038} [\href{https://arxiv.org/abs/1406.6351}{{\ttfamily 1406.6351}}].

\bibitem{Davison:2013jba}
R.~A. Davison, \emph{{Momentum relaxation in holographic massive gravity}},
  \href{https://doi.org/10.1103/PhysRevD.88.086003}{\emph{Phys. Rev.}
  {\bfseries D88} (2013) 086003}
  [\href{https://arxiv.org/abs/1306.5792}{{\ttfamily 1306.5792}}].

\bibitem{Davison:2014lua}
R.~A. Davison and B.~Gout\'eraux, \emph{{Momentum dissipation and effective
  theories of coherent and incoherent transport}},
  \href{https://doi.org/10.1007/JHEP01(2015)039}{\emph{JHEP} {\bfseries 01}
  (2015) 039} [\href{https://arxiv.org/abs/1411.1062}{{\ttfamily 1411.1062}}].

\bibitem{Davison:2015bea}
R.~A. Davison and B.~Gout\'eraux, \emph{{Dissecting holographic
  conductivities}}, \href{https://doi.org/10.1007/JHEP09(2015)090}{\emph{JHEP}
  {\bfseries 09} (2015) 090}
  [\href{https://arxiv.org/abs/1505.05092}{{\ttfamily 1505.05092}}].

\bibitem{Andrade:2018}
T.~Andrade and A.~Krikun, \emph{{Coherent vs incoherent transport in
  holographic strange insulators}},
  \href{https://doi.org/10.1007/JHEP05(2019)119}{\emph{JHEP} {\bfseries 05}
  (2019) 119} [\href{https://arxiv.org/abs/1812.08132}{{\ttfamily
  1812.08132}}].

\bibitem{Musso:2018wbv}
D.~Musso, \emph{{Simplest phonons and pseudo-phonons in field theory}},
  \href{https://arxiv.org/abs/1810.01799}{{\ttfamily 1810.01799}}.

\bibitem{Musso:2019kii}
D.~Musso and D.~Naegels, \emph{{Phonon and Shifton from a Real Modulated
  Scalar}},  \href{https://arxiv.org/abs/1907.04069}{{\ttfamily 1907.04069}}.

\bibitem{Delacretaz:2017zxd}
L.~V. Delacr\'etaz, B.~Gout\'eraux, S.~A. Hartnoll and A.~Karlsson,
  \emph{{Theory of hydrodynamic transport in fluctuating electronic charge
  density wave states}},
  \href{https://doi.org/10.1103/PhysRevB.96.195128}{\emph{Phys. Rev.}
  {\bfseries B96} (2017) 195128}
  [\href{https://arxiv.org/abs/1702.05104}{{\ttfamily 1702.05104}}].

\bibitem{Delacretaz:2016ivq}
L.~V. Delacr\'etaz, B.~Gout\'eraux, S.~A. Hartnoll and A.~Karlsson, \emph{{Bad
  Metals from Fluctuating Density Waves}},
  \href{https://doi.org/10.21468/SciPostPhys.3.3.025}{\emph{SciPost Phys.}
  {\bfseries 3} (2017) 025} [\href{https://arxiv.org/abs/1612.04381}{{\ttfamily
  1612.04381}}].

\bibitem{Ammon:2019wci}
M.~Ammon, M.~Baggioli and A.~Jimenez-Alba, \emph{{A Unified Description of
  Translational Symmetry Breaking in Holography}},
  \href{https://arxiv.org/abs/1904.05785}{{\ttfamily 1904.05785}}.

\bibitem{Baggioli:2019abx}
M.~Baggioli and S.~Grieninger, \emph{{Zoology of Solid \& Fluid Holography}},
  \href{https://arxiv.org/abs/1905.09488}{{\ttfamily 1905.09488}}.

\bibitem{Ooguri:2010kt}
H.~Ooguri and C.-S. Park, \emph{{Holographic End-Point of Spatially Modulated
  Phase Transition}},
  \href{https://doi.org/10.1103/PhysRevD.82.126001}{\emph{Phys. Rev.}
  {\bfseries D82} (2010) 126001}
  [\href{https://arxiv.org/abs/1007.3737}{{\ttfamily 1007.3737}}].

\bibitem{Donos:2011bh}
A.~Donos and J.~P. Gauntlett, \emph{{Holographic striped phases}},
  \href{https://doi.org/10.1007/JHEP08(2011)140}{\emph{JHEP} {\bfseries 08}
  (2011) 140} [\href{https://arxiv.org/abs/1106.2004}{{\ttfamily 1106.2004}}].

\bibitem{Donos:2013gda}
A.~Donos and J.~P. Gauntlett, \emph{{Holographic charge density waves}},
  \href{https://doi.org/10.1103/PhysRevD.87.126008}{\emph{Phys. Rev.}
  {\bfseries D87} (2013) 126008}
  [\href{https://arxiv.org/abs/1303.4398}{{\ttfamily 1303.4398}}].

\bibitem{Withers:2013loa}
B.~Withers, \emph{{Black branes dual to striped phases}},
  \href{https://doi.org/10.1088/0264-9381/30/15/155025}{\emph{Class. Quant.
  Grav.} {\bfseries 30} (2013) 155025}
  [\href{https://arxiv.org/abs/1304.0129}{{\ttfamily 1304.0129}}].

\bibitem{Withers:2014sja}
B.~Withers, \emph{{Holographic Checkerboards}},
  \href{https://doi.org/10.1007/JHEP09(2014)102}{\emph{JHEP} {\bfseries 09}
  (2014) 102} [\href{https://arxiv.org/abs/1407.1085}{{\ttfamily 1407.1085}}].

\bibitem{Ling:2014saa}
Y.~Ling, C.~Niu, J.~Wu, Z.~Xian and H.-b. Zhang, \emph{{Metal-insulator
  Transition by Holographic Charge Density Waves}},
  \href{https://doi.org/10.1103/PhysRevLett.113.091602}{\emph{Phys. Rev. Lett.}
  {\bfseries 113} (2014) 091602}
  [\href{https://arxiv.org/abs/1404.0777}{{\ttfamily 1404.0777}}].

\bibitem{Jokela:2014dba}
N.~Jokela, M.~Jarvinen and M.~Lippert, \emph{{Gravity dual of spin and charge
  density waves}}, \href{https://doi.org/10.1007/JHEP12(2014)083}{\emph{JHEP}
  {\bfseries 12} (2014) 083} [\href{https://arxiv.org/abs/1408.1397}{{\ttfamily
  1408.1397}}].

\bibitem{Cremonini:2016rbd}
S.~Cremonini, L.~Li and J.~Ren, \emph{{Holographic Pair and Charge Density
  Waves}}, \href{https://doi.org/10.1103/PhysRevD.95.041901}{\emph{Phys. Rev.}
  {\bfseries D95} (2017) 041901}
  [\href{https://arxiv.org/abs/1612.04385}{{\ttfamily 1612.04385}}].

\bibitem{Cai:2017qdz}
R.-G. Cai, L.~Li, Y.-Q. Wang and J.~Zaanen, \emph{{Intertwined order and
  holography: the case of the parity breaking pair density wave}},
  \href{https://doi.org/10.1103/PhysRevLett.119.181601}{\emph{Phys. Rev. Lett.}
  {\bfseries 119} (2017) 181601}
  [\href{https://arxiv.org/abs/1706.01470}{{\ttfamily 1706.01470}}].

\bibitem{Andrade:2017ghg}
T.~Andrade, A.~Krikun, K.~Schalm and J.~Zaanen, \emph{{Doping the holographic
  Mott insulator}},
  \href{https://doi.org/10.1038/s41567-018-0217-6}{\emph{Nature Phys.}
  {\bfseries 14} (2018) 1049}
  [\href{https://arxiv.org/abs/1710.05791}{{\ttfamily 1710.05791}}].

\bibitem{Donos:2012wi}
A.~Donos and J.~P. Gauntlett, \emph{{Black holes dual to helical current
  phases}}, \href{https://doi.org/10.1103/PhysRevD.86.064010}{\emph{Phys. Rev.}
  {\bfseries D86} (2012) 064010}
  [\href{https://arxiv.org/abs/1204.1734}{{\ttfamily 1204.1734}}].

\bibitem{Amoretti:2016bxs}
A.~Amoretti, D.~Areán, R.~Argurio, D.~Musso and L.~A. Pando~Zayas, \emph{{A
  holographic perspective on phonons and pseudo-phonons}},
  \href{https://doi.org/10.1007/JHEP05(2017)051}{\emph{JHEP} {\bfseries 05}
  (2017) 051} [\href{https://arxiv.org/abs/1611.09344}{{\ttfamily
  1611.09344}}].

\bibitem{Alberte:2017oqx}
L.~Alberte, M.~Ammon, A.~Jim\'enez-Alba, M.~Baggioli and O.~Pujolàs,
  \emph{{Holographic Phonons}},
  \href{https://doi.org/10.1103/PhysRevLett.120.171602}{\emph{Phys. Rev. Lett.}
  {\bfseries 120} (2018) 171602}
  [\href{https://arxiv.org/abs/1711.03100}{{\ttfamily 1711.03100}}].

\bibitem{Amoretti:2017frz}
A.~Amoretti, D.~Are\'an, B.~Gout\'eraux and D.~Musso, \emph{{Effective
  holographic theory of charge density waves}},
  \href{https://doi.org/10.1103/PhysRevD.97.086017}{\emph{Phys. Rev.}
  {\bfseries D97} (2018) 086017}
  [\href{https://arxiv.org/abs/1711.06610}{{\ttfamily 1711.06610}}].

\bibitem{Amoretti:2017axe}
A.~Amoretti, D.~Are\'an, B.~Gout\'eraux and D.~Musso, \emph{{DC resistivity of
  quantum critical, charge density wave states from gauge-gravity duality}},
  \href{https://doi.org/10.1103/PhysRevLett.120.171603}{\emph{Phys. Rev. Lett.}
  {\bfseries 120} (2018) 171603}
  [\href{https://arxiv.org/abs/1712.07994}{{\ttfamily 1712.07994}}].

\bibitem{Jokela:2017ltu}
N.~Jokela, M.~Jarvinen and M.~Lippert, \emph{{Pinning of holographic sliding
  stripes}}, \href{https://doi.org/10.1103/PhysRevD.96.106017}{\emph{Phys.
  Rev.} {\bfseries D96} (2017) 106017}
  [\href{https://arxiv.org/abs/1708.07837}{{\ttfamily 1708.07837}}].

\bibitem{Andrade:2017cnc}
T.~Andrade, M.~Baggioli, A.~Krikun and N.~Poovuttikul, \emph{{Pinning of
  longitudinal phonons in holographic spontaneous helices}},
  \href{https://doi.org/10.1007/JHEP02(2018)085}{\emph{JHEP} {\bfseries 02}
  (2018) 085} [\href{https://arxiv.org/abs/1708.08306}{{\ttfamily
  1708.08306}}].

\bibitem{Alberte:2017cch}
L.~Alberte, M.~Ammon, M.~Baggioli, A.~Jim\'enez and O.~Pujolàs, \emph{{Black
  hole elasticity and gapped transverse phonons in holography}},
  \href{https://doi.org/10.1007/JHEP01(2018)129}{\emph{JHEP} {\bfseries 01}
  (2018) 129} [\href{https://arxiv.org/abs/1708.08477}{{\ttfamily
  1708.08477}}].

\bibitem{Donos:2019tmo}
A.~Donos and C.~Pantelidou, \emph{{Holographic transport and density waves}},
  \href{https://doi.org/10.1007/JHEP05(2019)079}{\emph{JHEP} {\bfseries 05}
  (2019) 079} [\href{https://arxiv.org/abs/1903.05114}{{\ttfamily
  1903.05114}}].

\bibitem{Amoretti:2019cef}
A.~Amoretti, D.~Areán, B.~Goutéraux and D.~Musso, \emph{{Diffusion and
  universal relaxation of holographic phonons}},
  \href{https://doi.org/10.1007/JHEP10(2019)068}{\emph{JHEP} {\bfseries 10}
  (2019) 068} [\href{https://arxiv.org/abs/1904.11445}{{\ttfamily
  1904.11445}}].

\bibitem{Ammon:2019apj}
M.~Ammon, M.~Baggioli, S.~Gray and S.~Grieninger, \emph{{Longitudinal Sound and
  Diffusion in Holographic Massive Gravity}},
  \href{https://arxiv.org/abs/1905.09164}{{\ttfamily 1905.09164}}.

\bibitem{Donos:2019hpp}
A.~Donos, D.~Martin, C.~Pantelidou and V.~Ziogas, \emph{{Incoherent
  hydrodynamics and density waves}},
  \href{https://arxiv.org/abs/1906.03132}{{\ttfamily 1906.03132}}.

\bibitem{Armas:2019sbe}
J.~Armas and A.~Jain, \emph{{Viscoelastic hydrodynamics and holography}},
  \href{https://arxiv.org/abs/1908.01175}{{\ttfamily 1908.01175}}.

\bibitem{Donos:2019txg}
A.~Donos, D.~Martin, C.~Pantelidou and V.~Ziogas, \emph{{Hydrodynamics of
  broken global symmetries in the bulk}},
  \href{https://arxiv.org/abs/1905.00398}{{\ttfamily 1905.00398}}.

\bibitem{Donos:2013wia}
A.~Donos, \emph{{Striped phases from holography}},
  \href{https://doi.org/10.1007/JHEP05(2013)059}{\emph{JHEP} {\bfseries 05}
  (2013) 059} [\href{https://arxiv.org/abs/1303.7211}{{\ttfamily 1303.7211}}].

\bibitem{deHaro:2000vlm}
S.~de~Haro, S.~N. Solodukhin and K.~Skenderis, \emph{{Holographic
  reconstruction of space-time and renormalization in the AdS / CFT
  correspondence}}, \href{https://doi.org/10.1007/s002200100381}{\emph{Commun.
  Math. Phys.} {\bfseries 217} (2001) 595}
  [\href{https://arxiv.org/abs/hep-th/0002230}{{\ttfamily hep-th/0002230}}].

\bibitem{Kaminski:2009dh}
M.~Kaminski, K.~Landsteiner, J.~Mas, J.~P. Shock and J.~Tarrio,
  \emph{{Holographic Operator Mixing and Quasinormal Modes on the Brane}},
  \href{https://doi.org/10.1007/JHEP02(2010)021}{\emph{JHEP} {\bfseries 02}
  (2010) 021} [\href{https://arxiv.org/abs/0911.3610}{{\ttfamily 0911.3610}}].

\bibitem{Baggioli:2018bfa}
M.~Baggioli and A.~Buchel, \emph{{Holographic Viscoelastic Hydrodynamics}},
  \href{https://doi.org/10.1007/JHEP03(2019)146}{\emph{JHEP} {\bfseries 03}
  (2019) 146} [\href{https://arxiv.org/abs/1805.06756}{{\ttfamily
  1805.06756}}].

\bibitem{Withers:2018srf}
B.~Withers, \emph{{Short-lived modes from hydrodynamic dispersion relations}},
  \href{https://doi.org/10.1007/JHEP06(2018)059}{\emph{JHEP} {\bfseries 06}
  (2018) 059} [\href{https://arxiv.org/abs/1803.08058}{{\ttfamily
  1803.08058}}].

\bibitem{Grozdanov:2019kge}
S.~Grozdanov, P.~K. Kovtun, A.~O. Starinets and P.~Tadić, \emph{{Convergence
  of the Gradient Expansion in Hydrodynamics}},
  \href{https://doi.org/10.1103/PhysRevLett.122.251601}{\emph{Phys. Rev. Lett.}
  {\bfseries 122} (2019) 251601}
  [\href{https://arxiv.org/abs/1904.01018}{{\ttfamily 1904.01018}}].

\bibitem{Alberte:2015isw}
L.~Alberte, M.~Baggioli, A.~Khmelnitsky and O.~Pujolas, \emph{{Solid Holography
  and Massive Gravity}},
  \href{https://doi.org/10.1007/JHEP02(2016)114}{\emph{JHEP} {\bfseries 02}
  (2016) 114} [\href{https://arxiv.org/abs/1510.09089}{{\ttfamily
  1510.09089}}].

\bibitem{Hartnoll:2016tri}
S.~A. Hartnoll, D.~M. Ramirez and J.~E. Santos, \emph{{Entropy production,
  viscosity bounds and bumpy black holes}},
  \href{https://doi.org/10.1007/JHEP03(2016)170}{\emph{JHEP} {\bfseries 03}
  (2016) 170} [\href{https://arxiv.org/abs/1601.02757}{{\ttfamily
  1601.02757}}].

\bibitem{Alberte:2016xja}
L.~Alberte, M.~Baggioli and O.~Pujolas, \emph{{Viscosity bound violation in
  holographic solids and the viscoelastic response}},
  \href{https://doi.org/10.1007/JHEP07(2016)074}{\emph{JHEP} {\bfseries 07}
  (2016) 074} [\href{https://arxiv.org/abs/1601.03384}{{\ttfamily
  1601.03384}}].

\bibitem{Burikham:2016roo}
P.~Burikham and N.~Poovuttikul, \emph{{Shear viscosity in holography and
  effective theory of transport without translational symmetry}},
  \href{https://doi.org/10.1103/PhysRevD.94.106001}{\emph{Phys. Rev.}
  {\bfseries D94} (2016) 106001}
  [\href{https://arxiv.org/abs/1601.04624}{{\ttfamily 1601.04624}}].

\bibitem{Hartnoll:2012rj}
S.~A. Hartnoll and D.~M. Hofman, \emph{{Locally Critical Resistivities from
  Umklapp Scattering}},
  \href{https://doi.org/10.1103/PhysRevLett.108.241601}{\emph{Phys. Rev. Lett.}
  {\bfseries 108} (2012) 241601}
  [\href{https://arxiv.org/abs/1201.3917}{{\ttfamily 1201.3917}}].

\end{thebibliography}\endgroup
\end{document}